\documentclass[prd,amsmath,superscriptaddress,eqsecnum,twocolumn,nofootinbib]{revtex4-2}
\pdfoutput=1
\usepackage[utf8]{inputenc}

\usepackage{times}
\usepackage{microtype}
\usepackage{graphicx}
\usepackage{xcolor}
\usepackage[caption=false]{subfig} 
\usepackage{amssymb}
\usepackage{tensor}
\usepackage{adjustbox}
\usepackage{booktabs}
\usepackage{tabularx}
\usepackage{multirow}
\usepackage{enumitem}
\usepackage{etoolbox}
\usepackage{titlesec}

\def\sg{\textsl{g}}
\newcommand{\RN}[1]{\uppercase\expandafter{\romannumeral #1\relax}}

\makeatletter
\newcommand*{\defeq}{\mathrel{\rlap{%
			\raisebox{0.3ex}{$\m@th\cdot$}}%
		\raisebox{-0.3ex}{$\m@th\cdot$}}%
	=}
\newcommand*{\eqdef}{=\mathrel{\rlap{%
			\raisebox{0.3ex}{$\m@th\cdot$}}%
		\raisebox{-0.3ex}{$\m@th\cdot$}}%
}
\makeatother

\makeatletter
\g@addto@macro\bfseries{\boldmath}
\makeatother

\makeatletter
\def\thesubsection{\thesection.\arabic{subsection}}
\def\p@subsection{}
\makeatother

\titleformat{\section}[hang]
{\normalfont\normalsize\bfseries\MakeUppercase}
{\thesection.}{0.75em}
{\raggedright}

\titleformat{\subsection}[hang]
{\normalfont\normalsize\bfseries}
{\thesubsection.}{0.5em}
{\hspace{0pt}\raggedright}

\titleformat{\subsubsection}[hang]
{\normalfont\normalsize\bfseries}
{\thesubsubsection.}{1em}{\raggedright}

\titlespacing*{\section}{0pt}{5.25mm}{1mm}
\titlespacing*{\subsection}{0pt}{3mm}{0.5mm}
\titlespacing*{\subsubsection}{0pt}{2.25mm}{0.25mm}

\apptocmd{\appendix}{%
	\titleformat{\section}[hang]
	{\normalfont\normalsize\bfseries\MakeUppercase}
	{Appendix \thesection:}{0.5em}{\raggedright}%
	\titlespacing*{\section}{0pt}{4mm}{0.5mm}%
}{}{}

\apptocmd{\appendix}{%
	\titleformat{\subsection}[hang]
	{\normalfont\normalsize\bfseries}
	{\thesubsection.}{0.5em}{\raggedright}%
	\titlespacing*{\subsection}{0pt}{3mm}{0.4mm}%
}{}{}

\apptocmd{\appendix}{%
	\titleformat{\subsubsection}[hang]
	{\normalfont\normalsize\bfseries}
	{\thesubsubsection.}{0.5em}{\raggedright}%
	\titlespacing*{\subsubsection}{0pt}{2.5mm}{0.3mm}%
}{}{}

\usepackage{scalerel}
\usepackage{tikz}
\usetikzlibrary{svg.path}
\definecolor{orcidlogocol}{HTML}{A6CE39}
\tikzset{
	orcidlogo/.pic={
		\fill[orcidlogocol] svg{M256,128c0,70.7-57.3,128-128,128C57.3,256,0,198.7,0,128C0,57.3,57.3,0,128,0C198.7,0,256,57.3,256,128z};
		\fill[white] svg{M86.3,186.2H70.9V79.1h15.4v48.4V186.2z}
		svg{M108.9,79.1h41.6c39.6,0,57,28.3,57,53.6c0,27.5-21.5,53.6-56.8,53.6h-41.8V79.1z M124.3,172.4h24.5c34.9,0,42.9-26.5,42.9-39.7c0-21.5-13.7-39.7-43.7-39.7h-23.7V172.4z}
		svg{M88.7,56.8c0,5.5-4.5,10.1-10.1,10.1c-5.6,0-10.1-4.6-10.1-10.1c0-5.6,4.5-10.1,10.1-10.1C84.2,46.7,88.7,51.3,88.7,56.8z};
	}
}
\newcommand\orcidlink[1]{\href{https://orcid.org/#1}{\mbox{\scalerel*{
				\begin{tikzpicture}[yscale=-1,transform shape]
					\pic{orcidlogo};
				\end{tikzpicture}
			}{X}}}}
		
\usepackage{fancyhdr}

\usepackage{url,hyperref}
\hypersetup{colorlinks,linkcolor={blue!55!black},citecolor={red!50!black},urlcolor={blue!45!black},breaklinks=true}

\begin{document}
\title{Light rings and causality for nonsingular \\ ultracompact objects sourced by nonlinear electrodynamics}	
	
\author{Sebastian Murk\orcidlink{0000-0001-7296-0420}}
\email{sebastian.murk@oist.jp}
\affiliation{Quantum Gravity Unit, Okinawa Institute of Science and Technology, 1919-1 Tancha, Onna-son, Okinawa 904-0495, Japan}
	
\author{Ioannis Soranidis\orcidlink{0000-0002-8652-9874}}
\email{ioannis.soranidis@hdr.mq.edu.au}
\affiliation{School of Mathematical and Physical Sciences, Macquarie University, Sydney, New South Wales 2109, Australia}

\begin{abstract}
	We study observational signatures of nonsingular ultracompact objects regularized by nonlinear electrodynamics. The phenomenon of birefringence causes photons of different polarizations to propagate with respect to two distinct metrics, which manifests itself in the appearance of additional light rings surrounding the ultracompact object. We analyze the observational consequences of this result and illustrate our findings based on three regular black hole models commonly considered in the literature. We find that nonsingular horizonless ultracompact objects sourced by nonlinear electrodynamics possess an odd number of light rings and discuss the viability of this model as an effective description of their properties. In addition, we compare the phase velocities of polarized light rays propagating in nonsingular geometries sourced by nonlinear electrodynamics to the corresponding phase velocity in the Schwarzschild spacetime and demonstrate that regularizing the singularity by means of a theory that does not adhere to the Maxwell weak-field limit may lead to the emergence of acausal regions.
\end{abstract}

\maketitle
\tableofcontents
\thispagestyle{fancy}
		
\section{Introduction} \label{sec:introduction}
Nonlinear theories of electrodynamics were initially conceived with the intent to cure the divergences associated with the electric field self-energy of charged particles present in the linear Maxwell theory \cite{b:33,bi:34a,bi:34b}. Shortly after the 1933--1934 articles by Born and Infeld, Heisenberg and Euler devised a nonperturbative one-loop effective action that describes nonlinear corrections to Maxwell electrodynamics arising from quantum electrodynamical vacuum polarization effects (i.e., interactions of virtual electrons and positrons at the one-loop level without radiative corrections) \cite{he:36}. In addition to quantum field effects such as vacuum polarization, the exploration of nonlinear electrodynamics (NED) is also interesting from a general relativistic perspective since the Einstein field equations predict nonlinearities due to the gravitational coupling of electromagnetic fields \cite{p:61}. Elements of NED (most notably the nonlinear Born-Infeld action and its generalizations) also feature in different formulations of string theory/M-theory \cite{ft:85,l:89,t:97,cm:98,g:98,gh:01}, where in some instances NED appears as a low-energy effective field theory, which is at least partially responsible for its revival after a prolonged period of dormancy. Since then, possible applications of NED have grown significantly, reaching far beyond string theory.\footnote{Ref.~\cite{s:22} provides a short summary including recent developments.}\ Interestingly, NED theories coupled to gravity have proven extremely fruitful in the construction of new black hole solutions without singularities, so-called regular black holes (RBHs) \cite{pt:69,bmss:79,ag:98,ag:99a,ag:99b,b:00,b:01,bh:02,d:04,bv:14,fw:16,b:17,tsa:18,b:23}. This particular application of NED theories is the focus of the present article.
		
While the existence of dark massive ultracompact objects (UCOs) has been established beyond reasonable doubt, the question of whether or not the observed astrophysical black hole candidates possess defining black hole features such as singularities and horizons is still open \cite{f:14,cp:rev:19,mmt:rev:22,m:23}. This motivates the study of observational signatures (such as those presented in Tab.~\ref{tab:critical.lengths} and \ref{tab:light.rings} and of this article) that can distinguish between
\begin{enumerate}
	\item \hspace*{1mm} singular vs.\ nonsingular (``regular'')
	\item \hspace*{1mm} ``horizonful'' vs.\ horizonless 
\end{enumerate}
UCOs. In conjunction with gravitational wave detections and properties of accretion disks, optical signatures of the photon sphere such as light rings and shadows are among the principal observational tools expected to provide insights regarding the true physical nature of the observed astrophysical black hole candidates \cite{lgpv:18,cp:rev:19,v*:23,cdeh:24}. In this article, we analyze the physical properties of RBH solutions and closely related nonsingular horizonless UCO geometries arising from general relativity (GR) coupled to NED, focusing specifically on features that may lead to potentially measurable differences in their observational signatures. 	

The remainder of this article is organized as follows: In Sec.~\ref{sec:NED}, we briefly review the relevant properties of NED theories. In Sec.~\ref{sec:magn.sol}, we introduce the necessary ingredients needed to describe the purely magnetic NED solutions whose observational properties we investigate. In Sec.~\ref{sec:RBHs}, we summarize the properties of three regular UCO models commonly considered in the literature and derive explicit expressions for their respective effective geometries when the singularity resolution is achieved by means of NED. In Sec.~\ref{sec:LR}, we determine the location of light rings in both the background and the effective geometries of these models, discuss their dynamical behavior and stability, and compare characteristic observational features. In Sec.~\ref{sec:velocity}, we derive the phase velocity of photons moving in the effective geometry for different propagation directions and compare them to the Schwarzschild case. Lastly, in Sec.~\ref{sec:discussion.conclusions}, we summarize our results and outline their physical implications. For convenience, we review the phenomenon of birefringence in App.~\ref{sec:app:birefringence}. In App.~\ref{sec:app:singular.eff.geom}, we briefly comment on the singular behavior of the effective geometry. Throughout this article, we use the metric signature $(-,+,+,+)$ and work in dimensionless units such that $c=G=\hbar=k_{B}=1$.

\section{General relativity coupled to nonlinear electrodynamics} \label{sec:NED}
The electromagnetic field tensor is defined in terms of the electromagnetic four-vector potential $A_\mu$ as	
\begin{align}
	\tensor{F}{_\mu_\nu} \defeq \partial_\mu A_\nu - \partial_\nu A_\mu .
	\label{eq:def:Ftensor}
	\end{align}
Assuming a symmetric metric tensor $\tensor{\sg}{_\mu_\nu}$, only two independent algebraic invariants can be formed from an antisymmetric tensor $\tensor{F}{_\mu_\nu}$ \cite{p:61}, namely
\begin{align}
	\mathcal{F} = \tensor{F}{_\mu_\nu} \tensor{F}{^\mu^\nu} , \quad \mathcal{G} = \tensor{F}{_\mu_\nu}\! \left( \tensor[^\star]{F}{^\mu^\nu} \right) ,
	\label{eq:NEDinvariants}
\end{align}
where $\mathcal{F}$ denotes the electromagnetic field strength and $^\star$ the Hodge star operator, i.e., $\tensor[^\star]{F}{^\mu^\nu} \defeq \frac{1}{2} \tensor{\varepsilon}{^\mu^\nu^\rho^\sigma} \tensor{F}{_\rho_\sigma}\!$ with $\tensor{\varepsilon}{^\mu^\nu^\rho^\sigma}$ the Levi-Civita symbol. We restrict our considerations to theories in which the effective action involves a one-parameter Lagrangian density that is a local function of $\mathcal{F}$, i.e., $\mathcal{L}(\mathcal{F},\mathcal{G}) \equiv \mathcal{L}(\mathcal{F})$. In this case, the most general Lorentz-invariant action for GR coupled to NED in four dimensions is given by
\begin{align}
	S = \frac{1}{16 \pi} \int \sqrt{-\sg} \left[ R - \mathcal{L}(\mathcal{F}) \right] d^4x ,
	\label{eq:GR.NED.action}
\end{align}
where $\sg \equiv \det(\tensor{\sg}{_\mu_\nu})$ denotes the determinant of the metric tensor and $R = \tensor{\sg}{_\mu_\nu} \tensor{R}{^\mu^\nu}\!$ the Ricci curvature scalar.\footnote{Derivatives of $\tensor{F}{_\mu_\nu}\!$ are usually not considered in the action in order to avoid ghosts \cite{s:22}.}\ The generalized NED field equations obtained from the principle of least action and the Bianchi identities are given by
\begin{align} 
	\nabla_\mu \left( \mathcal{L}_\mathcal{F} \tensor{F}{^\mu^\nu} \right) = 0 , \quad \nabla_\mu \left( \tensor[^\star]{F}{^\mu^\nu} \right) = 0 ,
	\label{eq:NEDFE.BianchiIDs}
\end{align}
respectively, where $\mathcal{L}_\mathcal{F} \defeq \partial \mathcal{L} / \partial \mathcal{F}$ denotes the first-order derivative of the Lagrangian density with respect to the electromagnetic field strength. The coupled Einstein equations corresponding to the action of Eq.~\eqref{eq:GR.NED.action} are given by
\begin{align}
	\tensor{G}{_\mu_\nu} \defeq \tensor{R}{_\mu_\nu} - \frac{1}{2} R \tensor{\sg}{_\mu_\nu} = 8 \pi \tensor{T}{_\mu_\nu} , 
	\label{eq:EFE}
\end{align}
where the nonlinearities associated with the NED field enter via the energy-momentum tensor (EMT)
\begin{align}
	\tensor{T}{_\mu_\nu} = \frac{1}{4\pi} \left( \mathcal{L}_{\mathcal{F}}\tensor{F}{_\mu_\rho} \tensor{F}{^\rho_\nu} - \frac{1}{4}\tensor{\sg}{_\mu_\nu} \mathcal{L} \right) .
	\label{eq:EMT}
\end{align}
If the linear Maxwell theory is to be recovered in the weak-field limit (i.e., at small $\mathcal{F}$), then
\begin{align}
	\lim_{\mathcal{F} \to 0} \mathcal{L} \simeq \mathcal{F} , \quad \lim_{\mathcal{F} \to 0} \mathcal{L}_\mathcal{F} \simeq 1 ,
	\label{eq:weak-field.limit}
\end{align}
must hold. It is worth mentioning that there exists an alternative but formally equivalent description of NED theories based on a dual representation that is obtained by means of a Legendre transformation \cite{sgp:87}. However, for the purposes of our analysis in this article, it suffices to work in the Lagrangian formalism introduced above.
		
A static spherically symmetric metric is described by the line element
\begin{align}
	ds^2 = \tensor{\sg}{_\mu_\nu} dx^{\mu} dx^{\nu} = - f(r) dt^2 + f(r)^{-1} dr^2 + r^2 d\Omega^2 , 
	\label{eq:metric}
\end{align}
where $r$ denotes the areal radius and $d\Omega^2$ the normalized spherically symmetric Riemannian metric on the 2-sphere $\mathbb{S}^2$, which is given in terms of angular coordinates $(\theta,\phi)$ by $d\Omega^2 \equiv d\theta^2 + \sin^2 \theta \; \! d\phi^2$. In spherical symmetry, the only nonvanishing components of the electromagnetic field tensor are $\tensor{F}{_t_r} = - \tensor{F}{_r_t}\!$ (corresponding to a radial electric field) and $\tensor{F}{_\theta_\phi} = - \tensor{F}{_\phi_\theta}\!$ (corresponding to a radial magnetic field). From Eq.~\eqref{eq:NEDFE.BianchiIDs}, the electric and magnetic charge are then identified as
\begin{align}
	Q_e = - r^2 \mathcal{L}_\mathcal{F} \tensor{F}{_t_r} , \quad 
	Q_m = - \frac{\tensor{F}{_\theta_\phi}}{\sin \theta} ,
\end{align}
respectively. One of the properties that make NED coupled to gravity an appealing candidate field theory for the construction of nonsingular spacetimes is that the EMT associated with the gauge-invariant Lagrangian density $\mathcal{L}(\mathcal{F})$ [cf.\ Eq.~\eqref{eq:EMT}] describes the spherically symmetric vacuum due to the symmetries $\tensor{T}{_t^t} = \tensor{T}{_r^r}$ (invariance under boosts in the radial direction) and $\tensor{T}{_\theta^\theta} = \tensor{T}{_\phi^\phi}$ (spherical symmetry) \cite{d:92}.
		
With the exception of Maxwell and Born-Infeld theories, a salient feature of NED theories is birefringence \cite{bb:70,ndsk:00,lkns:00}, i.e., the phenomenon that the two photon modes/polarizations associated with the two degrees of freedom encoded in the field effectively propagate with respect to two distinct metrics (see App.~\ref{sec:app:birefringence} for details). In the case $\mathcal{L} = \mathcal{L}(\mathcal{F})$ that we consider, one of them coincides with the background metric $\tensor{\sg}{_\mu_\nu}$ of Eq.~\eqref{eq:metric} which solves the field equations of GR coupled to NED [Eq.~\eqref{eq:EFE} with the EMT of Eq.~\eqref{eq:EMT}], and the other --- referred to as effective metric in what follows --- is given by
\begin{align}
	\tensor{\bar{\sg}}{^\mu^\nu} = \tensor{\sg}{^\mu^\nu} - 4 \frac{\mathcal{L}_{\mathcal{F}\mathcal{F}}}{\mathcal{L}_\mathcal{F}} \tensor{F}{^\mu_\rho} \tensor{F}{^\rho^\nu} ,
	\label{eq:def.eff.metric}
\end{align}
with the deviation from the background metric arising from the nonlinearity of the electromagnetic field, where $\mathcal{L}_{\mathcal{F} \mathcal{F}} \defeq \partial \mathcal{L}_\mathcal{F} / \partial \mathcal{F}$ denotes the second-order derivative of $\mathcal{L}(\mathcal{F})$ with respect to the field strength $\mathcal{F}$, and the bar label is used to distinguish physical quantities associated with the effective geometry from those of the background geometry here and in what follows. In the linear Maxwell theory $\tensor{\bar{\sg}}{^\mu^\nu} \equiv \tensor{\sg}{^\mu^\nu}$ due to $\mathcal{L}_{\mathcal{F}} = \mathrm{const.}$ and $\mathcal{L}_{\mathcal{F}\mathcal{F}}=0$, and thus there is no birefringence.
		
Lastly, we note that solutions with an electric charge $Q_e \neq 0$ (including dyonic solutions where both $Q_e \neq 0$ and $Q_m \neq 0$) generally require a non-Maxwell behavior of the Lagrangian density $\mathcal{L}(\mathcal{F})$ at small $\mathcal{F}$ in order to maintain a regular center \cite{bmss:79,b:01,b:17}, i.e., in this case the Lagrangian density does not conform to the weak-field limit of Eq.~\eqref{eq:weak-field.limit}. Although purely electric solutions ($Q_e \neq 0$ and $Q_m=0$) with a regular center have been proposed in Refs.~\cite{ag:98,ag:99a,ag:99b}, any such solution inevitably requires a Lagrangian density that behaves nonlocally in the sense that different functions $\mathcal{L}(\mathcal{F})$ are required in different spacetime domains (specifically at small and large $r$ with the former behaving non-Maxwellian in the limit $\mathcal{F} \to 0$) as pointed out in Refs.~\cite{b:00,b:01}.\footnote{We also note the proposal of hybrid solutions which can circumvent the no-go theorem for solutions with an electric charge \cite{bmss:79} by means of a phase transition to a dual magnetic phase near the core such that the electric field does not extend all the way to the center of the solution \cite{bh:02}.}\ We therefore restrict our considerations to purely magnetic solutions ($Q_e=0$ and $Q_m \neq 0$) in what follows. Prototypical examples of such solutions are the RBH models proposed by Bardeen \cite{b:68} and Hayward \cite{h:06}.

\section{Magnetic solutions in nonlinear electrodynamics} \label{sec:magn.sol}
As motivated in the previous section, we restrict our considerations to purely magnetic solutions, i.e., $Q_e=0$ and $Q_m \neq 0$ (cf.\ Sec.~III in Ref.~\cite{fw:16}). In this case, the generic form of the Lagrangian density that solves the Einstein equations of GR coupled to NED [Eqs.~\eqref{eq:EFE}--\eqref{eq:EMT}] in spherical symmetry [Eq.~\eqref{eq:metric}] is given by
\begin{align}
	\mathcal{L}(r) = - 2 \left( \frac{f'(r)}{r} + \frac{f(r)-1}{r^2} \right) . 
\end{align}
Using the identification $f(r) \equiv 1 - 2m(r)/r$, one further obtains
\begin{align}
	\mathcal{L}(r) = \frac{4m'(r)}{r^2} .
\end{align}
The four-potential, magnetic charge, and electromagnetic field strength are given by 
\begin{align}
	A_{\mu} = \left(0,0,0,Q_{m}\cos{\theta}\right) , \quad Q_{m} = \frac{q^2}{\sqrt{2\alpha}} , \quad \mathcal{F} = \frac{2 Q^2_m}{r^4} ,
	\label{eq:A.Qm.F}
\end{align}
respectively, where the parameter $\alpha>0$ has dimensions of length squared and $q$ denotes a free integration constant with dimensions of length. The precise form of the Lagrangian density depends on the choice of the mass function $m(r)$. We proceed with the generic form proposed by Fan and Wang (cf.\ Eq.~(26) in Ref.~\cite{fw:16}), namely
\begin{align}
	\mathcal{L}(\mathcal{F}) = \frac{4 \mu (\alpha \mathcal{F})^{\frac{\nu+3}{4}}}{\alpha \left[1+(\alpha \mathcal{F})^{\frac{\nu}{4}}\right]^{1+\frac{\mu}{\nu}}} ,
	\label{eq:L.gen}
\end{align}
where $\mu,\nu>0$ are dimensionless constants and the choices $(\mu,\nu)=(3,2)$, $(\mu,\nu)=(3,3)$, and $(\mu,\nu)=(3,1)$ correspond to the Bardeen \cite{b:68}, Hayward \cite{h:06}, and the Maxwellian \cite{fw:16} RBH solutions, respectively. The geometry is specified in terms of $\mu$, $\nu$, $\alpha$, and $q$ by the metric function
\begin{align}
	f(r) = 1 - \frac{2 M r^{\mu-1}}{\left(r^{\nu}+q^{\nu}\right)^{\mu/\nu}} ,
	\label{eq:f.gen}
\end{align}
where $M \equiv q^3/\alpha$ corresponds to the gravitational mass (which in this case coincides with the electromagnetically induced mass, cf.\ Eqs.~(8)--(9) of Ref.~\cite{tsa:18} and the discussion therein). In spherical symmetry, the effective metric according to which photons with one of the two possible polarizations move along null geodesics is given by [cf.\ Eq.~\eqref{eq:def.eff.metric}]\footnote{The covariant effective metric tensor components $\tensor{\bar{\sg}}{_\mu_\nu}$ are identified via the relation $\tensor{\bar{\sg}}{^\mu^\rho}\tensor{\bar{\sg}}{_\rho_\nu}=\tensor{\delta}{^\mu_{\!\! \nu}}$.}\
\begin{align}
	\tensor{\bar{\sg}}{_t_t} &= \tensor{\sg}{_t_t}, \quad \tensor{\bar{\sg}}{_r_r} = \tensor{\sg}{_r_r}, \label{eq:eff.metric.tt.rr} \\ 
	\tensor{\bar{\sg}}{_\theta_\theta} &= \tensor{\sg}{_\theta_\theta} \left( 1 + \frac{2 \mathcal{F} \mathcal{L}_{\mathcal{F}\mathcal{F}}}{\mathcal{L}_{\mathcal{F}}} \right)^{-1}, \label{eq:eff.metric.thetatheta} \\ 
	\tensor{\bar{\sg}}{_\phi_\phi} &= \tensor{\sg}{_\phi_\phi} \left( 1 + \frac{2 \mathcal{F} \mathcal{L}_{\mathcal{F}\mathcal{F}}}{\mathcal{L}_{\mathcal{F}}} \right)^{-1} ,
	\label{eq:eff.metric.phiphi}
\end{align}
where the first line follows from $\tensor{F}{_t_r} \propto Q_e =0$ as we consider purely magnetic solutions with no electric charge. Since $\tensor{F}{_\theta_\phi} \propto Q_m$ and $\mathcal{F} \propto Q_m^2$ [cf.\ Eq.~\eqref{eq:A.Qm.F}], one would naively expect that in the absence of magnetic charge ($Q_m=0$) the angular effective metric tensor components of Eqs.~\eqref{eq:eff.metric.thetatheta} and \eqref{eq:eff.metric.phiphi} reduce to those of the background metric as well, that is $\tensor{\bar{\sg}}{_\theta_\theta} = \tensor{\sg}{_\theta_\theta}$ and $\tensor{\bar{\sg}}{_\phi_\phi} = \tensor{\sg}{_\phi_\phi}$. However, this is not always the case. This peculiar behavior is linked to the weak-field limit of the corresponding NED Lagrangian density $\mathcal{L}(\mathcal{F})$. As we shall see in what follows, the quantity $\mathcal{F}\mathcal{L}_{\mathcal{F}\mathcal{F}}/\mathcal{L}_{\mathcal{F}}$ does not vanish in the limit $Q_m\rightarrow 0$ in some RBH models.

\section{Regular black holes and nonsingular horizonless ultracompact objects} \label{sec:RBHs}
While keeping our analysis generic, we illustrate our results based on the RBH models proposed by Bardeen \cite{b:68}, Hayward \cite{h:06}, and Cadoni \textit{et al.}\ \cite{c*:23} in what follows. These models are characterized by the strength of their respective deformations from the Schwarzschild geometry (with the model by Cadoni \textit{et al.}\ corresponding to the strongest possible deformation) and exhibit different weak-field limits. Using the generic form of the Lagrangian density introduced in Eq.~\eqref{eq:L.gen} of the previous section [or, equivalently, the generic form of the metric function $f(r)$ in Eq.~\eqref{eq:f.gen}], each model is described by a distinct choice of the parameters $\alpha>0$, $q>0$, and $\nu>0$.
		
We note here that in the original formulation of nonsingular RBH geometries the regularization of spacetime is achieved by means of a minimal length scale $\ell$ that appears in the metric function [cf.\ Eqs.~\eqref{eq:f.B}, \eqref{eq:f.H}, and \eqref{eq:f.C}] and acts as a Planckian cutoff beyond which GR is no longer valid. An alternative interpretation is to consider these metrics as a solution of the Einstein field equations with an EMT sourced by NED, cf.\ Eqs.~\eqref{eq:EFE}--\eqref{eq:EMT} \cite{b:01,fw:16,b:17,tsa:18}. In the purely magnetic case that we consider due to the arguments laid out in the last paragraph of Sec.~\ref{sec:NED}, these types of solutions describe the gravitational field of a magnetic monopole. The difference in the interpretation of $\ell$ as a Planckian cutoff vs.\ $\ell$ arising from NED effects is that in the latter case macroscopic values are permissible [see Sec.~\ref{subsec:LR.loc}, Tab.~\ref{tab:critical.lengths}]. Other ways of regularizing the black hole spacetime singularity have also been proposed in the literature, e.g., in 4D Einstein-Gauß-Bonnet theories \cite{nn:23}, loop quantum gravity \cite{aos:23}, and higher-dimensional approaches \cite{bch:24}. As we shall see in Sec.~\ref{sec:LR}, the precise quantification and comparison of light ring signatures may allow us to distinguish between nonsingular geometries obtained via different regularization methods and ultimately identify the underlying effective theory describing regular UCOs.

Depending on the minimal length scale $\ell$ and mass $M$, the geometry specified by the metric of Eq.~\eqref{eq:metric} with $f(r)$ given by Eq.~\eqref{eq:f.gen} can represent different types of UCOs. Setting $M \equiv 1$ and focusing on the case where the roots are real and positive [$r \in \mathbb{R}_{>0}$], solving the equation $f(r)=0$ may result in either of the following distinct outcomes:
\begin{enumerate}[label=\Roman*.]
	\item 
	\begin{tabbing}
		\quad \= Two roots [$\ell<\ell_c$], which corresponds to \\
		\> an RBH with an inner and an outer horizon;
	\end{tabbing}
	\item 
	\begin{tabbing}
		\quad \= One root [$\ell = \ell_c$], which corresponds to \\
		\> an extremal RBH with one degenerate horizon;
	\end{tabbing}
	\item 	
	\begin{tabbing}
		\quad \= No roots [$\ell>\ell_c$], which corresponds to \\
		\> a nonsingular horizonless UCO;
	\end{tabbing}
\end{enumerate}
Here, $\ell_c$ denotes the critical length at which the outer and inner horizon coalesce, $r_+(\ell_c) \equiv r_-(\ell_c)$.
	
\subsection{Bardeen model} \label{subsec:B.RBH}
The Bardeen RBH \cite{b:68} is described by the metric function 
\begin{align}
	f_{\mathcal{B}}(r) = 1 - \frac{2 M r^2}{\left( r^2 + \ell^2 \right)^{3/2}} ,
	\label{eq:f.B}
\end{align}
where $\ell$ denotes the aforementioned minimal length scale introduced to regularize the black hole spacetime. Comparison with the generic metric function of Eq.~\eqref{eq:f.gen} identifies the coefficients as $\mu_\mathcal{B}=3$, $\nu_\mathcal{B}=2$, and we can rewrite Eq.~\eqref{eq:f.B} as
\begin{align}
	f_{\mathcal{B}}(r) = 1 - \frac{2 q^3 r^2}{\alpha (r^2+q^2)^{3/2}} .
	\label{eq:RBH.B.f}
\end{align}
This implies $q = \ell$ (note that $q \propto \ell$ is always expected based on our argumentation in the previous section) and [using Eq.~\eqref{eq:A.Qm.F}]
\begin{align}
	\alpha = \frac{\ell^3}{M} , \quad Q_{m} = \sqrt{\frac{M \ell}{2}} .
	\label{eq:B.RBH.alpha.Qm}
\end{align} 
The NED Lagrangian density is given by
\begin{align}
	\mathcal{L}_{\mathcal{B}}(\mathcal{F}) = \frac{12 (\alpha\mathcal{F})^{5/4}}{\alpha \left[ 1+(\alpha\mathcal{F})^{1/2} \right]^{5/2}},
	\label{eq:B.RBH.L}
\end{align}
and expanding about the point $\mathcal{F}=0$ reveals that the weak-field behavior of the NED Bardeen RBH is $\sim \mathcal{O}(\mathcal{F}^{5/4})$, i.e., stronger than that of the linear Maxwell theory. 
	
Solving the equation $f_{\mathcal{B}}(r)=0$ allows us to determine the roots representing the locations of the outer $r^{(\mathcal{B})}_{+}$ and inner $r^{(\mathcal{B})}_{-}$ horizon. Since the exact expressions are quite lengthy, we provide the leading-order terms in the series expansion of small $\ell$ instead:
\begin{align}
	r^{(\mathcal{B})}_{+} &= 2M - \frac{3 \ell^2}{4M} + \mathcal{O}(\ell^4) , \\
	r^{(\mathcal{B})}_{-} &= \frac{\ell^{3/2}}{\sqrt{2M}} + \frac{3 \ell^{5/2}}{8 \sqrt{2} M^{3/2}} + \mathcal{O} \big(\ell^{7/2} \big) .
\end{align} 
The critical length at which the two horizons coincide describes an extremal black hole. Solving $r^{(\mathcal{B})}_{+}=r^{(\mathcal{B})}_{-}$ leads to 
\begin{align}
	\ell^{(\mathcal{B})}_{c} = \frac{4M}{3\sqrt{3}} .
	\label{eq:B.RBH.lc}
\end{align}
For the Bardeen NED Lagrangian density Eq.~\eqref{eq:B.RBH.L}, the effective metric tensor components are given explicitly by 
\begin{align}
	\tensor{\bar{\sg}}{_t_t} &= - f_{\mathcal{B}}(r), \quad \tensor{\bar{\sg}}{_r_r} = f_{\mathcal{B}}(r)^{-1}, 
	\label{eq:B.RBH.eff.metric.tt.rr}  \\ 
	\tensor{\bar{\sg}}{_\theta_\theta} &= \tensor{\sg}{_\theta_\theta} \left( \frac{3}{2} -\frac{7\ell^2}{2(r^2+\ell^2)} \right)^{-1}, 
	\label{eq:B.RBH.eff.metric.thetatheta} \\ 
	\tensor{\bar{\sg}}{_\phi_\phi} &= \tensor{\sg}{_\phi_\phi}  \left( \frac{3}{2} -\frac{7\ell^2}{2(r^2+\ell^2)} \right)^{-1} ,
	\label{eq:B.RBH.eff.metric.phiphi} 
\end{align}
where we observe that for vanishing magnetic charge $Q_m=0$ [or, equivalently, vanishing minimal length $\ell=0$ by virtue of Eq.~\eqref{eq:B.RBH.alpha.Qm}], the angular effective metric components do not coincide with those of the background metric tensor. The same behavior is exhibited in the asymptotic limit $r\rightarrow\infty$. The individual components of the term $\mathcal{F} \mathcal{L}_{\mathcal{F}\mathcal{F}} / \mathcal{L}_{\mathcal{F}}$ conspire in such a way so as to not yield the correct asymptotic behavior, which is an unavoidable byproduct of the non-Maxwellian weak-field limit behavior. 
	
\subsection{Hayward model} \label{subsec:H.RBH}
The Hayward RBH \cite{h:06} is described by the metric function 
\begin{align}
	f_{\mathcal{H}}(r) = 1 - \frac{2 M r^2}{r^3 + 2 M \ell^2} .
	\label{eq:f.H}
\end{align}
Comparison with the generic metric function of Eq.~\eqref{eq:f.gen} identifies the coefficients as $\mu_\mathcal{H}=3$, $\nu_\mathcal{H}=3$, and we can rewrite Eq.~\eqref{eq:f.H} as
\begin{align}
	f_{\mathcal{H}}(r) = 1 - \frac{2 q^3r^2}{\alpha (r^3+q^3)}.
	\label{eq:RBH.H.f}
\end{align}
This implies $q = (2 M \ell^2)^{1/3}$ and [again using Eq.~\eqref{eq:A.Qm.F}]
\begin{align}
	\alpha = 2 \ell^2, \quad Q_{m} = \frac{M^{2/3} \ell^{1/3}}{2^{1/3}} .
	\label{eq:H.RBH.alpha.Qm} 
\end{align} 
The NED Lagrangian density is given by
\begin{align}
	\mathcal{L}_{\mathcal{H}}(\mathcal{F}) = \frac{12(\alpha\mathcal{F})^{3/2}}{\alpha \left[1+(\alpha \mathcal{F})^{3/4} \right]^2},
	\label{eq:H.RBH.L}
\end{align}
and expanding about the point $\mathcal{F}=0$ reveals that the weak-field behavior of the NED Hayward RBH is $\sim \mathcal{O}(\mathcal{F}^{3/2})$, which is stronger compared to both the linear Maxwell theory and the NED Bardeen RBH discussed in Sec.~\ref{subsec:B.RBH}. 
	
Analogous to the previous subsection, we solve the equation $f_{\mathcal{H}}(r)=0$ to determine the roots corresponding to the outer $r^{(\mathcal{H})}_{+}$ and inner $r^{(\mathcal{H})}_{-}$ horizon locations of the Hayward model. Once again, the exact expressions are rather lengthy, and thus we provide the leading terms in their series expansions for small $\ell$ instead:
\begin{align}
	r^{(\mathcal{H})}_{+} &= 2M - \frac{\ell^2}{2M} + \mathcal{O}(\ell^4), \\ 
	r^{(\mathcal{H})}_{-} &= \ell + \frac{\ell^{2}}{4M} + \mathcal{O}(\ell^{3}).
\end{align} 
The critical length at which the two horizons coincide and the Hayward RBH becomes extremal is given by
\begin{align}
	r^{(\mathcal{H})}_{+} = r^{(\mathcal{H})}_{-} \quad \Rightarrow \quad	\ell^{(\mathcal{H})}_{c} = \frac{4M}{3\sqrt{3}} ,
	\label{eq:H.RBH.lc}
\end{align}
which coincides with the expression obtained for the Bardeen RBH [cf.\ Eq.~\eqref{eq:B.RBH.lc}]. For the Hayward NED Lagrangian density Eq.~\eqref{eq:H.RBH.L}, the effective metric tensor components are given explicitly by	
\begin{align}
	\tensor{\bar{\sg}}{_t_t} &= - f_{\mathcal{H}}(r), \quad \tensor{\bar{\sg}}{_r_r} = f_{\mathcal{H}}(r)^{-1}, 
	\label{eq:H.RBH.eff.metric.tt.rr} \\ 
	\tensor{\bar{\sg}}{_\theta_\theta} &= \tensor{\sg}{_\theta_\theta} \left(2 -\frac{9M\ell^2}{r^3+2M\ell^2} \right)^{-1},
	\label{eq:H.RBH.eff.metric.thetatheta} \\ 
	\tensor{\bar{\sg}}{_\phi_\phi} &= \tensor{\sg}{_\phi_\phi}  \left(2 -\frac{9M\ell^2}{r^3+2M\ell^2} \right)^{-1}.
	\label{eq:H.RBH.eff.metric.phiphi} 
\end{align}
As in the Bardeen model, the angular effective metric tensor components do not exhibit the correct limiting behaviors, neither for vanishing magnetic charge $Q_m =0$ [$\ell=0$] nor in the asymptotic regime $r\to\infty$. In contrast to the Bardeen case where the relation between $Q_m$ and $M$ [cf.\ Eq.~\eqref{eq:B.RBH.alpha.Qm}] leads to exact cancelations in the $\mathcal{F} \mathcal{L}_{\mathcal{F}\mathcal{F}} / \mathcal{L}_{\mathcal{F}}$ term, the effective metric tensor components arising from the Hayward NED Lagrangian density [Eq.~\eqref{eq:H.RBH.L}] depend explicitly on the gravitational mass $M$.
	
\subsection{Cadoni \textit{et al.}\ model} \label{subsec:C.RBH}
The model considered by Cadoni \textit{et al.}\ in Ref.~\cite{c*:23} belongs to the Maxwellian class of solutions first described in Sec.~III.C.\ of Ref.~\cite{fw:16}. Its metric function is given by
\begin{align}
	f_{\mathcal{C}}(r) = 1 - \frac{2Mr^2}{(r+\ell)^3} .
	\label{eq:f.C}
\end{align}
Comparison with the generic metric function of Eq.~\eqref{eq:f.gen} identifies the coefficients as $\mu_\mathcal{C}=3$, $\nu_\mathcal{C}=1$, and Eq.~\eqref{eq:f.C} can be rewritten as
\begin{align}
	f_{\mathcal{C}}(r) = 1 - \frac{2 q^3 r^2}{\alpha (r+q)^3} .
	\label{eq:RBH.C.f}
\end{align}
This implies $q = \ell$ and
\begin{align}
	\alpha = \frac{\ell^3}{M}, \quad Q_{m} = \sqrt{\frac{M \ell}{2}} ,
	\label{eq:C.RBH.alpha.Qm}
\end{align} 
analogous to the relations obtained for the Bardeen RBH described in Sec.~\ref{subsec:B.RBH} [cf.\ Eq.~\eqref{eq:B.RBH.alpha.Qm}]. The NED Lagrangian density is given by
\begin{align}
	\mathcal{L}_{\mathcal{C}}(\mathcal{F}) = \frac{12 \mathcal{F}}{\left[ 1 + (\alpha \mathcal{F})^{1/4} \right]^4} ,
	\label{eq:C.RBH.L}
\end{align}
and expanding about the point $\mathcal{F}=0$ reveals that the weak-field behavior of the NED Cadoni \textit{et al.}\ RBH is $\sim\mathcal{O}(\mathcal{F})$, i.e., it reduces to the linear Maxwell theory.
	
Analogous to the previous two subsections, we solve the equation $f_{\mathcal{C}}(r)=0$ to determine the roots corresponding to the outer $r^{(\mathcal{C})}_{+}$ and inner $r^{(\mathcal{C})}_{-}$ horizon locations of the Cadoni \textit{et al.}\ model. Once again, the exact expressions are rather lengthy, and thus we provide the leading terms in their series expansions for small $\ell$ instead:
\begin{align}
	r^{(\mathcal{C})}_{+} &= 2M - 3\ell - \frac{3\ell^2}{2M} + \mathcal{O}(\ell^3), \\ 
	r^{(\mathcal{C})}_{-} &= \frac{\ell^{3/2}}{\sqrt{2M}} + \frac{3\ell^{2}}{4M} + \mathcal{O}(\ell^{5/2}).
\end{align} 
The critical length at which the two horizons coincide and the RBH considered by Cadoni \textit{et al.}\ becomes extremal is given by
\begin{align}
	r^{(\mathcal{C})}_{+} = r^{(\mathcal{C})}_{-} \quad \Rightarrow \quad	\ell^{(\mathcal{C})}_{c} = \frac{8M}{27} ,
	\label{eq:C.RBH.lc}
\end{align}
which is smaller compared to the critical lengths of the Bardeen and Hayward RBH $\ell^{(\mathcal{B})}_{c} = \ell^{(\mathcal{H})}_{c} = 4M/(3\sqrt{3})$ [cf.\ Eqs.~\eqref{eq:B.RBH.lc} and \eqref{eq:H.RBH.lc}].
	
For the Cadoni \textit{et al.}\ NED Lagrangian density Eq.~\eqref{eq:C.RBH.L}, the effective metric tensor components are given explicitly by	
\begin{align}
	\tensor{\bar{\sg}}{_t_t} &= - f_{\mathcal{C}}(r), \quad \tensor{\bar{\sg}}{_r_r} = f_{\mathcal{C}}(r)^{-1}, 
	\label{eq:C.RBH.eff.metric.tt.rr} \\ 
	\tensor{\bar{\sg}}{_\theta_\theta} &= \tensor{\sg}{_\theta_\theta} \left(1 - \frac{5\ell}{2 \left( r + \ell \right)} \right)^{-1}, 
	\label{eq:C.RBH.eff.metric.thetatheta} \\ 
	\tensor{\bar{\sg}}{_\phi_\phi} &= \tensor{\sg}{_\phi_\phi}  \left(1 -\frac{5\ell}{2\left( r + \ell \right)} \right)^{-1}.
	\label{eq:C.RBH.eff.metric.phiphi} 
\end{align}
It is evident from these relations that the effective metric tensor components of the NED Cadoni \textit{et al.}\ RBH admit the correct limiting behaviors for vanishing minimal length and in the asymptotic regime, in contrast to those of the Bardeen [cf.\ Eqs.~\eqref{eq:B.RBH.eff.metric.tt.rr}--\eqref{eq:B.RBH.eff.metric.phiphi}] and Hayward [cf.\ Eqs.~\eqref{eq:H.RBH.eff.metric.tt.rr}--\eqref{eq:H.RBH.eff.metric.phiphi}] RBH models. This desirable property is inherently linked to the Maxwellian behavior $\sim \mathcal{O}(\mathcal{F})$ of the Cadoni \textit{et al.}\ NED Lagrangian density [Eq.~\eqref{eq:C.RBH.L}] for small field strengths $\mathcal{F}$. Since the relation between $Q_m$ and $M$ in the model considered by Cadoni \textit{et al.}\ is the same as that of the Bardeen RBH [cf.\ Eqs.~\eqref{eq:B.RBH.alpha.Qm} and \eqref{eq:C.RBH.alpha.Qm}], the effective metric tensor components have once again no explicit dependence on the gravitational mass $M$, unlike those of the Hayward RBH.

\section{Light rings} \label{sec:LR}
\subsection{Mathematical prerequisites} \label{subsec:LR.mp}
The Lagrangian for the motion of a free particle in a curved spacetime is given by 
\begin{align}
	L_{p} = \frac{1}{2} \tensor{\sg}{_\mu_\nu} \dot{x}^{\mu} \dot{x}^{\nu} ,
\end{align} 
where the dot denotes a derivative with respect to an appropriately chosen affine parameter (e.g., the proper time $\tau$ for timelike geodesics) characterizing the trajectory. Due to the spherical symmetry imposed by Eq.~\eqref{eq:metric}, we can limit our considerations to trajectories in the equatorial plane [$\theta=\pi/2$, $\dot{\theta}=0$] without loss of generality in what follows. Denoting the four-velocity $u^\mu=(\dot{t},\dot{r},0,\dot{\phi})$, we have
\begin{align}
	L_{p} = \frac{u^2}{2} = \frac{1}{2} \left( \tensor{\sg}{_t_t} \dot{t}^2 + \tensor{\sg}{_r_r} \dot{r}^2 + \tensor{\sg}{_\phi_\phi} \dot{\phi}^2 \right),
\end{align}
where $u^2 \defeq u_\mu u^\mu$. Using the corresponding Euler-Lagrange equations, it is straightforward to confirm that $t$ and $\phi$ are cyclic variables,\footnote{Recall that the metric of Eq.~\eqref{eq:metric} is spherically symmetric and static, and thus $\tensor{\sg}{_\mu_\nu}$ is independent of these coordinates.}\ resulting by virtue of Noether's theorem \cite{n:1918} in the conservation of their associated conjugate variables, namely the energy $E$ and angular momentum $L$, i.e.,
\begin{align}
	\frac{\partial L_{p}}{\partial \dot{t}} = \tensor{\sg}{_t_t} \dot{t} = - E, \quad \frac{\partial L_{p}}{\partial \dot{\phi}} = \tensor{\sg}{_\phi_\phi}\dot{\phi} = L . 
	\label{eq:cons.E.L}
\end{align}
As our interest lies in studying light rings formed by the null geodesics of photons, we substitute $\dot{t}$ and $\dot{\phi}$ from Eq.~\eqref{eq:cons.E.L} into
\begin{align}
	\tensor{\sg}{_\mu_\nu}\dot{x}^{\mu}\dot{x}^{\nu} = 0
	\label{eq:bkg.nc}
\end{align}
to obtain
\begin{align}
	\dot{r}^2 + V(r) = 0, \quad V(r) \defeq \frac{E^2}{\tensor{\sg}{_t_t} \tensor{\sg}{_r_r}} + \frac{L^2}{\tensor{\sg}{_r_r} \tensor{\sg}{_\phi_\phi}} .
	\label{eq:V}
\end{align}
The location of the light rings is determined by the conditions $\dot{r}=0$ and $\ddot{r}=0$, which imply $V(r)=0$ and $V'(r)=0$, respectively \cite{cbh:17,ch:20}. Solving these equations for $r$ yields the radius of the light ring and the impact parameter $b \defeq L/E$. 
	
As alluded to previously (see Sec.~\ref{sec:NED}), UCOs sourced by NED will in general (i.e., with the exception of Maxwell and Born-Infeld theories) exhibit birefringence, and thus the two possible photon polarization propagate with respect to two different metrics and at different velocities.\footnote{It is worth noting that Lagrangian densities based on the Born-Infeld nonlinearity cannot give rise to nonsingular black hole solutions in spherically symmetric settings \cite{hi:37}.}\ Analogous to the derivation for the background metric above, photons propagating on null geodesics with respect to the effective metric [cf.\ Eqs.~\eqref{eq:def.eff.metric} and \eqref{eq:eff.metric.tt.rr}--\eqref{eq:eff.metric.phiphi}] satisfy
\begin{align}
	\tensor{\bar{\sg}}{_\mu_\nu} \dot{x}^{\mu} \dot{x}^{\nu} = 0,
	\label{eq:eff.nc}
\end{align} 
and we obtain
\begin{align}
	\dot{r}^2 + \bar{V}(r) = 0, \quad \bar{V}(r) \defeq \frac{\bar{E}^2}{\tensor{\bar{\sg}}{_t_t} \tensor{\bar{\sg}}{_r_r}} + \frac{\bar{L}^2}{\tensor{\bar{\sg}}{_r_r} \tensor{\bar{\sg}}{_\phi_\phi}} ,
	\label{eq:Vbar}
\end{align}
with the light ring locations equivalently specified by the conditions $\bar{V}(r)=0$ and $\bar{V}'(r)=0$. An alternative method to identify the light ring locations is to determine the critical points of the function (see Ref.~\cite{cbh:17} for a detailed derivation)
\begin{align}
	H = \frac{-\tensor{\sg}{_t_\phi}\pm \sqrt{\tensor{\sg}{_t_\phi}-\tensor{\sg}{_t_t}\tensor{\sg}{_\phi_\phi}}}{\tensor{\sg}{_\phi_\phi}} \stackrel{\eqref{eq:metric}}{=} \frac{\sqrt{-\tensor{\sg}{_t_t}\tensor{\sg}{_\phi_\phi}}}{\tensor{\sg}{_\phi_\phi}} ,
	\label{eq:H}
\end{align}		
which simplifies to the rightmost expression for static spherically symmetric metrics of the form of Eq.~\eqref{eq:metric}. The advantage of this approach is that it implicitly incorporates the condition $V(r)=0$, thereby allowing one to disregard the conserved conjugate variables $E$ and $L$. The same line of thought applies for the effective metric. We now proceed with the identification of light ring locations for the three RBH models described in Sec.~\ref{sec:RBHs}.
	
\subsection{Light ring locations} \label{subsec:LR.loc}
\begin{figure*}[!htbp]
	\vspace*{-6mm}
	\subfloat[{Bardeen model [Sec.~\ref{subsec:B.RBH}, Eq.~\eqref{eq:B.RBH.L}]}]{
		\includegraphics[width=.485\linewidth]{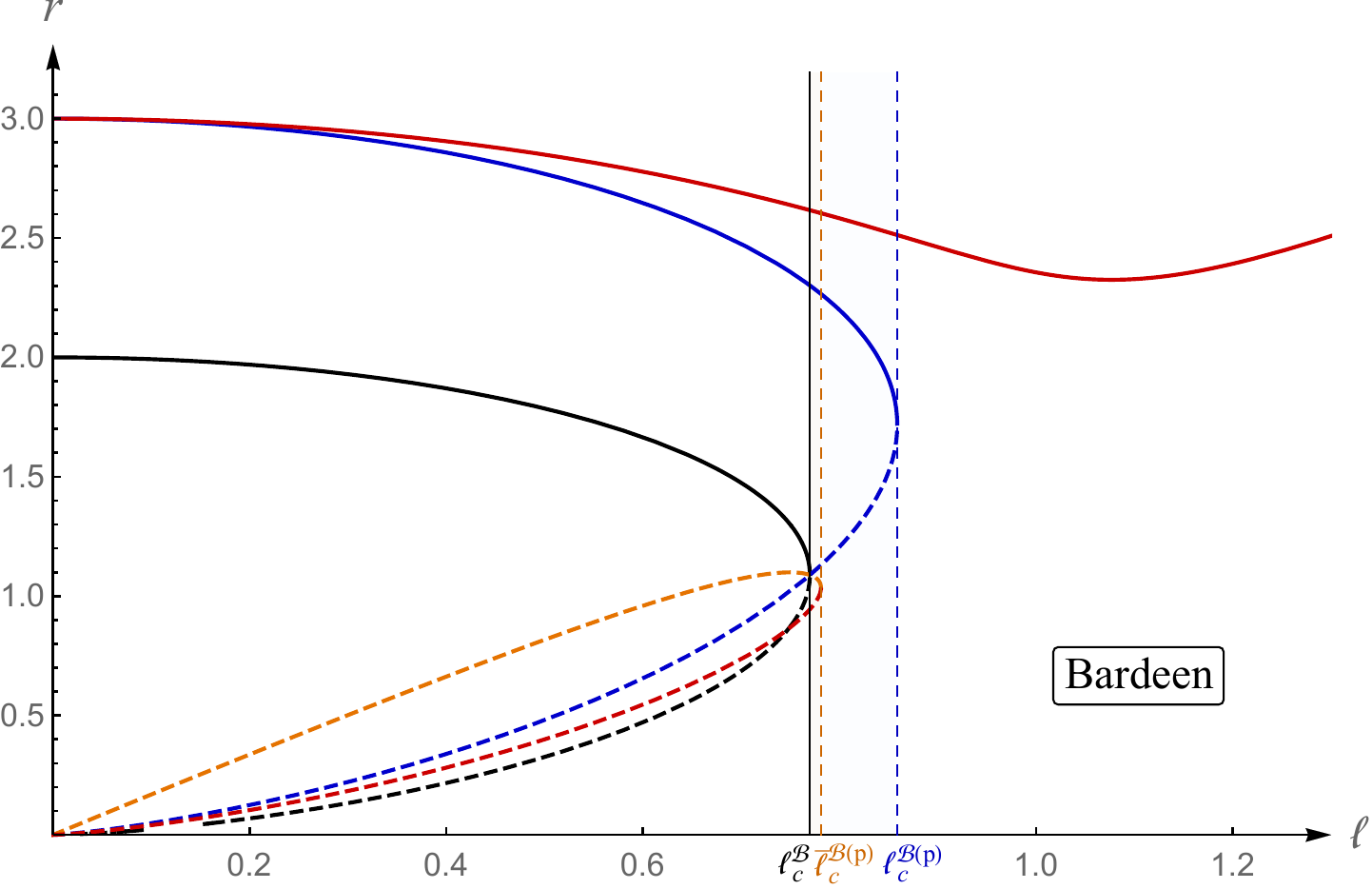}
		\label{subfig:B.RBH}
	}\hfill
	\subfloat[{Hayward model [Sec.~\ref{subsec:H.RBH}, Eq.~\eqref{eq:H.RBH.L}]}]{
		\includegraphics[width=.485\linewidth]{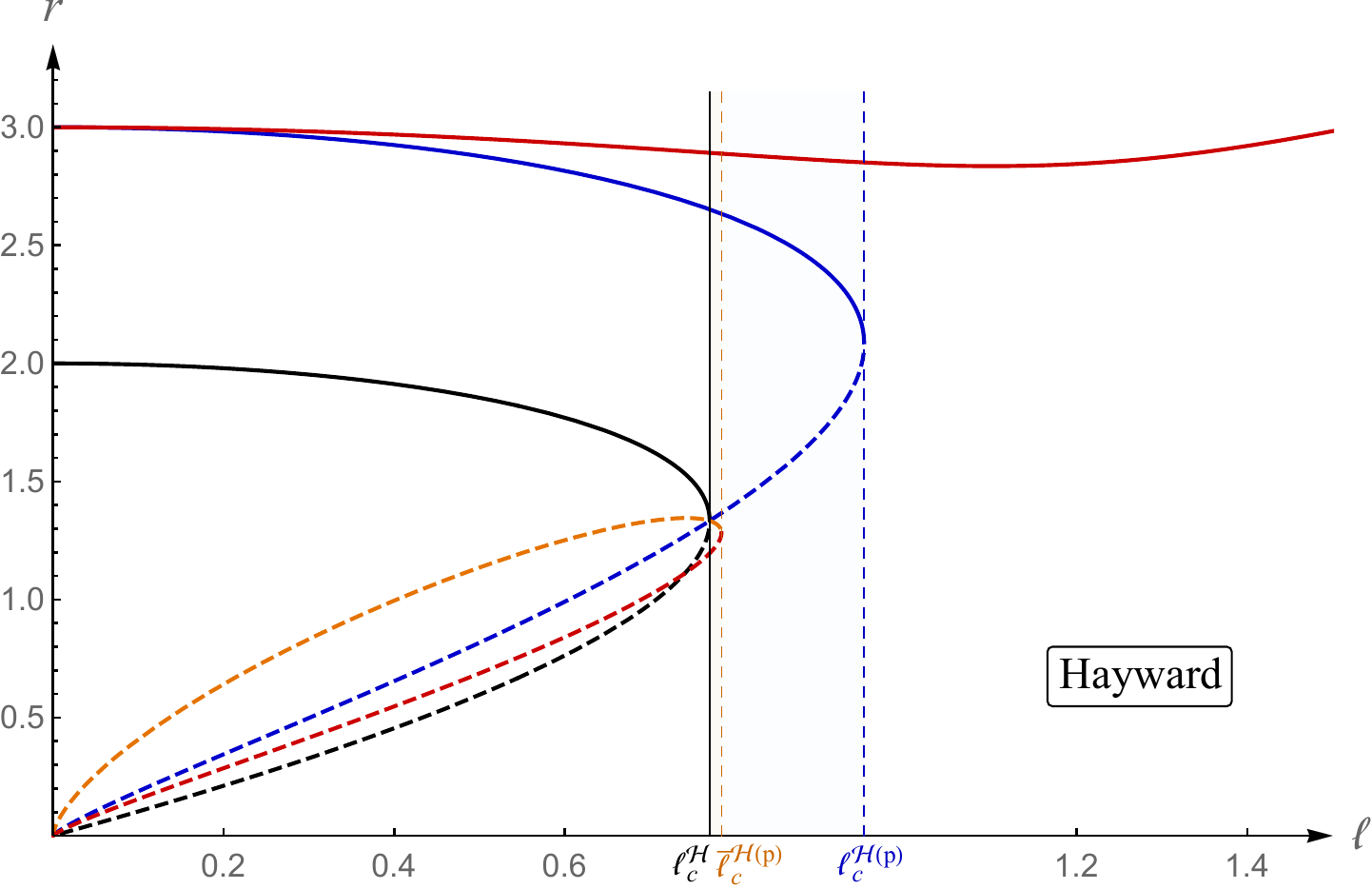}
		\label{subfig:H.RBH}
	}\vspace*{-2mm}\
	\subfloat[{Cadoni \textit{et al.}\ model [Sec.~\ref{subsec:C.RBH}, Eq.~\eqref{eq:C.RBH.L}]}]{
		\includegraphics[width=.485\linewidth]{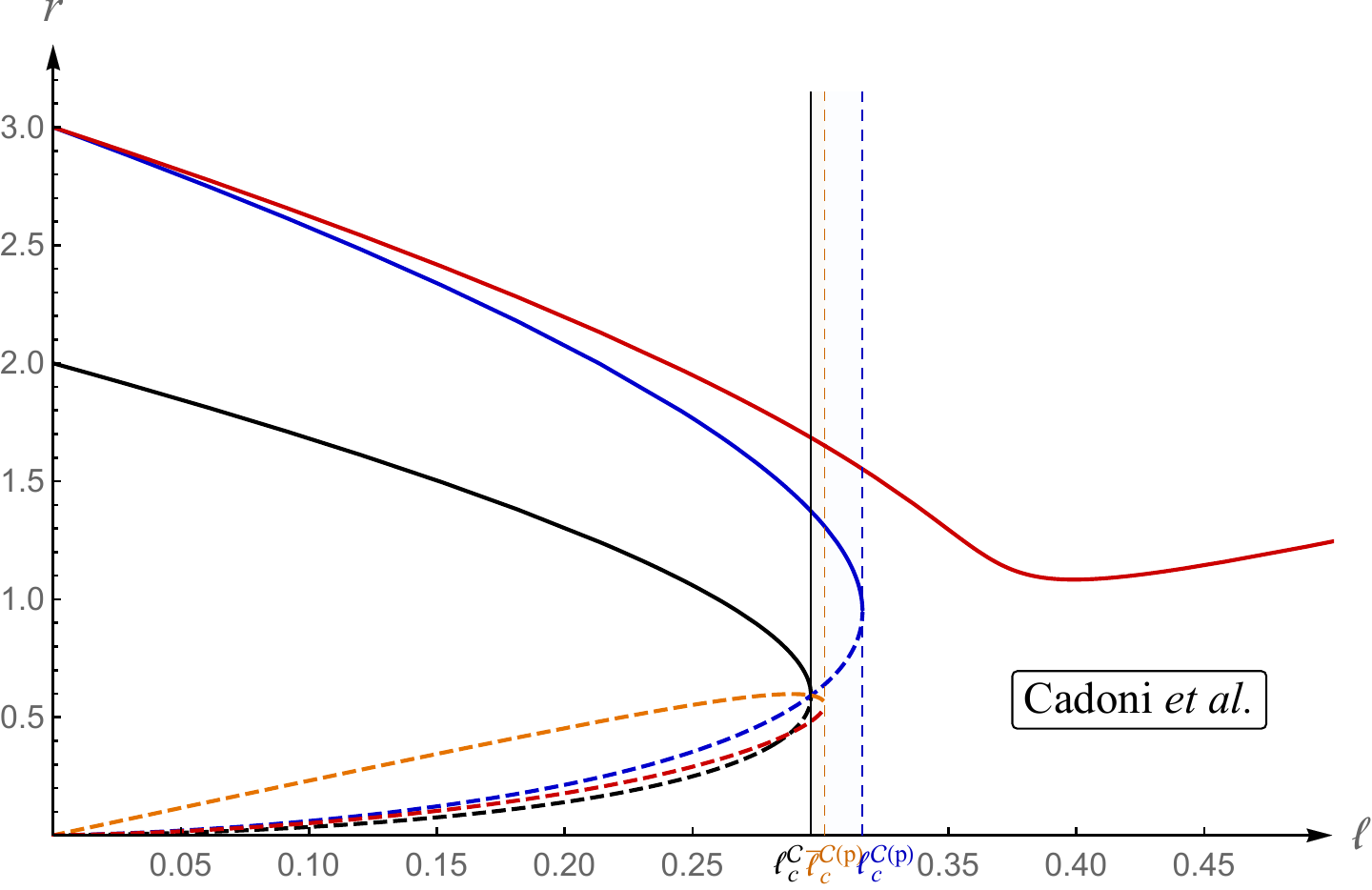}
		\label{subfig:C.RBH}
	}\hfill
	\subfloat[{Bardeen (gray) vs.\ Hayward (blue) vs.\ Cadoni \textit{et al.}\ (red)}]{
		\includegraphics[width=.485\linewidth]{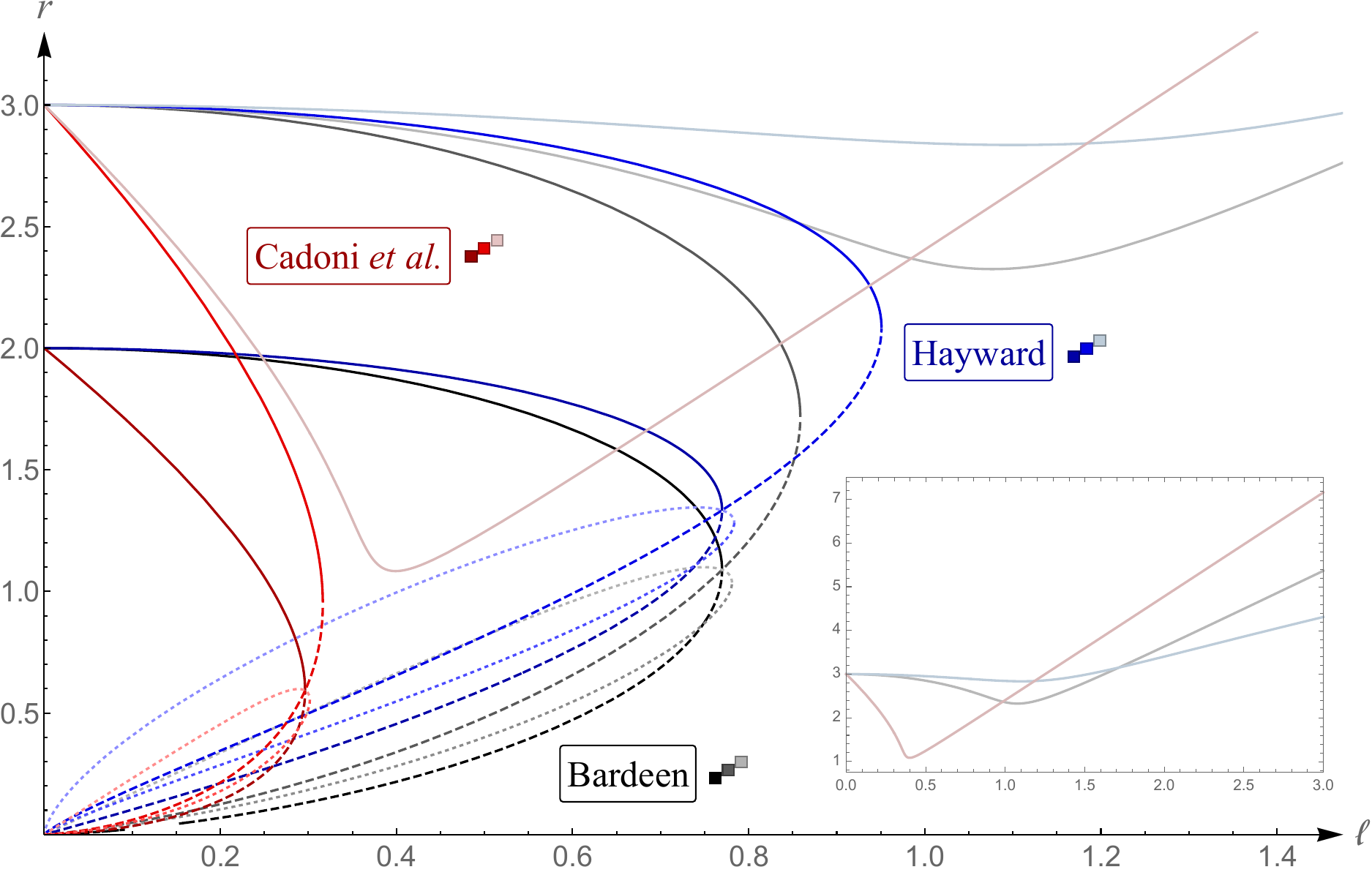}
		\label{subfig:LR.comparison}
	}
	\caption{Horizons, critical lengths, and light rings for the three nonsingular UCO models discussed in Sec.~\ref{sec:RBHs}. In Subfigs.~\ref{subfig:B.RBH}--\ref{subfig:C.RBH}, the solid (dashed) black line represents the outer (inner) horizon $r_+$ ($r_-$), the solid (dashed) blue line represents the outer (inner) light ring $r^+_p$ ($r^-_p$) in the background geometry [Eq.~\eqref{eq:metric}], and the solid (dashed) red line represents the outer (inner) light ring $\bar{r}^+_p$ ($\bar{r}^-_p$) in the effective geometry [Eq.~\eqref{eq:def.eff.metric}], which harbors a third light ring $\bar{r}^0_p$ situated between the two whose location is indicated by the dashed orange line. The critical lengths $\ell_c$, $\bar{\ell}_c^{(p)}$, and $\ell_c^{(p)}$ at which the horizons [Eqs.~\eqref{eq:B.RBH.lc} \eqref{eq:H.RBH.lc}, \eqref{eq:C.RBH.lc}], the two innermost light rings in the effective geometry, and the inner and outer light ring in the background geometry merge are indicated by the thin vertical black, thin vertical dashed orange, and thin vertical dashed blue line, respectively. The regularization parameter domain $0<\ell\leqslant\ell_{c}$ ($\ell>\ell_c$) describes an RBH (a nonsingular horizonless UCO). The minimal length interval $\ell_{c}<\ell<\bar{\ell}^{(p)}_{c}$ ($\bar{\ell}^{(p)}_{c}<\ell<\ell^{(p)}_{c}$) is indicated by the region shaded in light orange (light blue) and corresponds to the second (third) column in Tab.~\ref{tab:light.rings}. The outer light ring in the effective geometry of the Bardeen (Hayward) [Cadoni \textit{et al.}] model has a global minimum at $(r,\bar{\ell}^{+(\mathcal{B})}_\text{min})=(2.3251,1.0766)$ $\big( (r,\bar{\ell}^{+(\mathcal{H})}_\text{min})=(2.8355,1.1004) \big)$ $\big[ (r,\bar{\ell}^{+(\mathcal{C})}_\text{min})=(1.0834,0.3991) \big]$. In Fig.~\ref{subfig:LR.comparison}, the Bardeen (Hayward) [Cadoni \textit{et al.}] model is represented by the gray (blue) [red] color scheme. In each of the color schemes, the outer (inner) horizons are represented by the solid (dashed) line in the darkest hue, the outer (inner) light ring in the background geometry by the solid (dashed) line in the medium hue, the two innermost light rings in the effective geometry by the dotted lines in the two lightest hues, and the outer light ring in the effective geometry by the solid line in the lightest hue. Differences in the critical horizon and light ring length scales of the three models (see also Tab.~\ref{tab:critical.lengths}) are attributable to their different deformation strengths from the Schwarzschild geometry. The inset in the bottom right-hand side corner of Fig.~\ref{subfig:LR.comparison} serves as a comparison of the outer light rings in the effective geometry of the three models, illustrating their characteristic behavior for varying minimal length $\ell$ in the interval $0 < \ell \leqslant 3$.} 
	\label{fig:Horizons.LRs}
	\vspace*{-2mm}
\end{figure*}

The methodology of our calculation can be applied analogously for the three RBH models described in Sec.~\ref{sec:RBHs} [and any other RBH geometry whose metric can be cast into a form analogous to Eqs.~\eqref{eq:RBH.B.f}, \eqref{eq:RBH.H.f}, and \eqref{eq:RBH.C.f}]. For the three models considered in Sec.~\ref{sec:RBHs}, the explicit form of the potentials in Eqs.~\eqref{eq:V} and \eqref{eq:Vbar} are given by
\begin{align}
	V_\mathcal{B}(r) &= -E^2 + L^2 \left( \frac{1}{r^2} - \frac{2M}{\left(r^2 + \ell^2\right)^{3/2}} \right) , \\
	V_\mathcal{H}(r) &= -E^2 + L^2 \left( \frac{1}{r^2} - \frac{2M}{r^3 + 2 M \ell^2} \right) , \\
	V_\mathcal{C}(r) &= - E^2 + L^2 \left(\frac{1}{r^2} - \frac{2M}{\left(r+\ell\right)^3}\right),
\end{align} 
and
\begin{align} 
	\bar{V}_\mathcal{B}(r) &= -E^2 + \frac{L^2 \left(3r^2-4\ell^2\right)\left((r^2+\ell^2)^{3/2}-2Mr^2\right)}{2r^2 (r^2+\ell^2)^{5/2}} , \\
	\bar{V}_\mathcal{H}(r) &= -E^2 + \frac{L^2 \left(2r^3-5M\ell^2\right)\left(r^3 +2M(\ell^2-r^2)\right)}{\left(r^4+2M\ell^2\right)^2} , \\
	\bar{V}_\mathcal{C}(r) &= - E^2 + \frac{L^2 (2r-3\ell) \left((r+\ell)^3-2Mr^2\right)}{2r^2 \left(r+\ell\right)^4} , 
\end{align}
respectively. The light ring locations in the background and in the effective geometry are determined by solving the equations $V'(r)=0$ and $\bar{V}'(r)=0$ for $r$. In what follows, we use a superscript ``$+$'' (``$-$'') to label the outer (inner) light ring $r^{+}_{p}$ ($r^{-}_{p}$) and a subscript ``$p$'' for ``photosphere'' to unambiguously distinguish the locations of the light rings from those of the outer and inner horizon $r_{+}$ and $r_{-}$, respectively. Once again, many of the exact expressions are quite lengthy and somewhat cumbersome to deal with by hand, and thus we do not provide them here. For the interested reader, all explicit expressions are provided in the \textsc{GitHub} repository linked as Ref.~\cite{m:github}.\footnote{The code in this repository is written in \textsc{Mathematica 12} \cite{Mathematica12}.}\ Since the procedure is analogous for the background and the effective geometry and follows the same steps for each RBH model, we do not repeat them explicitly here. The inner and outer horizons as well as the inner and outer light rings in the background [Eq.~\eqref{eq:metric}] and effective [Eq.~\eqref{eq:def.eff.metric}] geometry for the three UCO models considered in Sec.~\ref{sec:RBHs} are illustrated in Fig.~\ref{fig:Horizons.LRs} for the mass parameter $M = 1$.

For illustrative purposes, we focus on the model by Cadoni \textit{et al.}\ in what follows since it is the only one out of the three models that exhibits the correct limiting behaviors (see last paragraph of Sec.~\ref{subsec:C.RBH}). The leading terms in the series expansions of the outer and inner light ring in the background geometry of this model about the point $\ell=0$ are given by
\begin{align}
	r^{+}_{p} &= 3M - 4 \ell - \frac{2 \ell^2}{M} + \mathcal{O}(\ell^3), \label{eq:C.RBH.oLR.bkg} \\ 
	r^{-}_{p} &= \frac{\ell^{4/3}}{(3M)^{1/3}} + \frac{4 \ell^{5/3}}{3^{5/3} M^{2/3}} + \frac{2 \ell^2}{3M} + \mathcal{O}(\ell^{7/3}) ,
	\label{eq:C.RBH.iLR.bkg}
\end{align}
respectively. Based on these expressions, we can ascertain that the outer light ring is situated closer to the black hole than in the Schwarzschild geometry, where it is located at $r=3M$. As one would expect, the limit of vanishing minimal length $\lim_{\ell \to 0} r^{+}_{p}=3M$ reduces to this expression. The inner light ring on the other hand vanishes in this limit, $\lim_{\ell \to 0} r^{-}_{p}=0$. The models proposed by Bardeen and Hayward exhibit the same qualitative behavior, which can be verified by examining their exact expressions and/or confirmed graphically as in Fig.~\ref{fig:Horizons.LRs}. Similar to the critical length $\ell_c$ at which the inner and outer horizon merge, there is a critical length $\ell_c^{(p)}$, $r^{+}_{p}\big(\ell_c^{(p)}\big) \equiv r^{-}_{p}\big(\ell_c^{(p)}\big)$, at which the inner and outer light rings merge (and then disappear) in the background geometry. This is in line with the analyses of Refs.~\cite{cdlv:23,flv:24} which examine properties of the EMT and quasinormal mode spectra for geometries that smoothly interpolate between RBHs and nonsingular horizonless UCOs based on the value of the regularization parameter $\ell$. However, the light ring signature in the effective geometry significantly differs from that of the background geometry. As depicted in Fig.~\ref{fig:Horizons.LRs}, there are generally three distinct light rings in the effective geometry. The outermost light ring (indicated by the solid red line in Figs.~\ref{subfig:B.RBH}--\ref{subfig:C.RBH}) persists and never disappears (in contrast to the behavior of the outer light ring in the background geometry indicated by the solid blue line), whereas the middle (dashed orange line) and innermost (dashed red line) light rings are present up to some critical length $\bar{\ell}^{(p)}_{c}$ indicated by the thin vertical dashed orange line in Figs.~\ref{subfig:B.RBH}--\ref{subfig:C.RBH}. These characteristics are consistent across all three UCO models. Interestingly, the radius of the outer light ring in the effective geometry gradually decreases with increasing minimal length $\ell$ until it reaches a global minimum value at some $\bar{\ell}^+_\text{min} > \ell_c^{(p)}$ from which point onwards its radius starts to continuously increase with increasing $\ell$. We also note that for each model there is a small minimal length scale interval $\ell_c < \ell < \bar{\ell}_c^{(p)}$ (indicated by the region shaded in light orange that is enclosed by the thin vertical solid black line signifying $\ell_c$ and the thin vertical dashed orange line signifying $\bar{\ell}_c^{(p)}$ in Figs.~\ref{subfig:B.RBH}--\ref{subfig:C.RBH}) in which the two innermost light rings of the effective geometry are no longer obscured by the horizons (as the UCO is no longer an RBH but a nonsingular horizonless UCO at length scales $\ell > \ell_c$) and become visible to external observers. Similarly, there is a small minimal length interval $\ell_c < \ell < \ell_c^{(p)}$ (indicated by the union of the aforementioned interval $\ell_c < \ell < \bar{\ell}_c^{(p)}$ shaded in light orange and the region shaded in light blue that is enclosed by the thin vertical dashed orange line signifying $\bar{\ell}_c^{(p)}$ and the thin vertical dashed blue line signifying $\ell_c^{(p)}$ in Figs.~\ref{subfig:B.RBH}--\ref{subfig:C.RBH}), in which the inner light of the background geometry becomes visible.
	
\begin{figure*}[!htbp]
	\vspace*{-1mm}
	\subfloat[{Outer LR difference $\bar{r}^{+}_{p} - r^{+}_{p}$ for the Bardeen model [Sec.~\ref{subsec:B.RBH}, Eq.~\eqref{eq:B.RBH.L}]}]{
		\includegraphics[width=.485\linewidth]{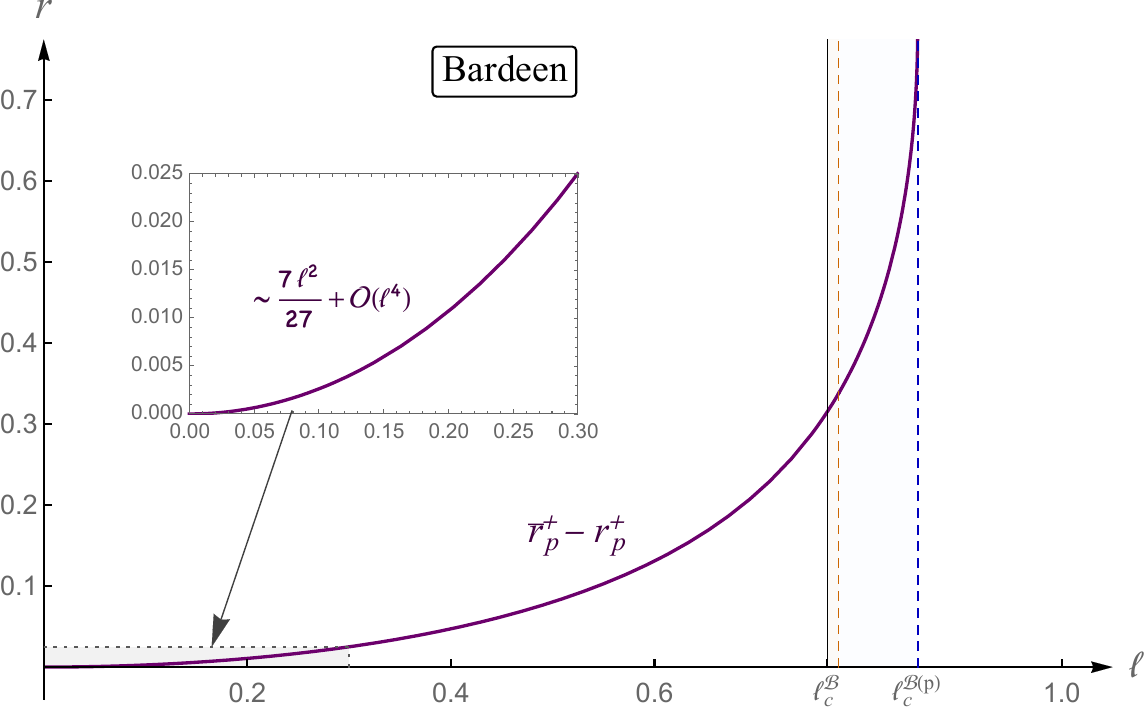}
		\label{subfig:B.Diff}
	}\hfill
	\subfloat[{Outer LR difference $\bar{r}^{+}_{p} - r^{+}_{p}$ for the Hayward model [Sec.~\ref{subsec:H.RBH}, Eq.~\eqref{eq:H.RBH.L}]}]{
		\includegraphics[width=.485\linewidth]{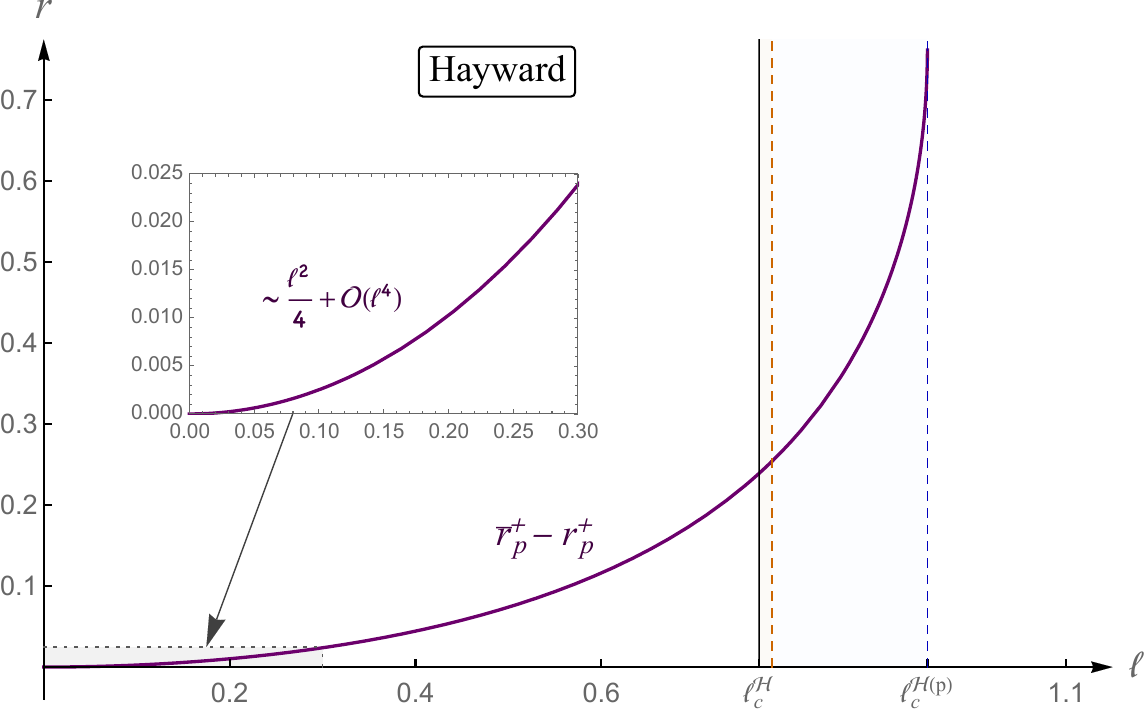}
		\label{subfig:H.Diff}
	}\
	\subfloat[{Outer LR difference $\bar{r}^{+}_{p} - r^{+}_{p}$ for the Cadoni \textit{et al.} model [Sec.~\ref{subsec:C.RBH}, Eq.~\eqref{eq:C.RBH.L}]}]{
		\includegraphics[width=.505\linewidth]{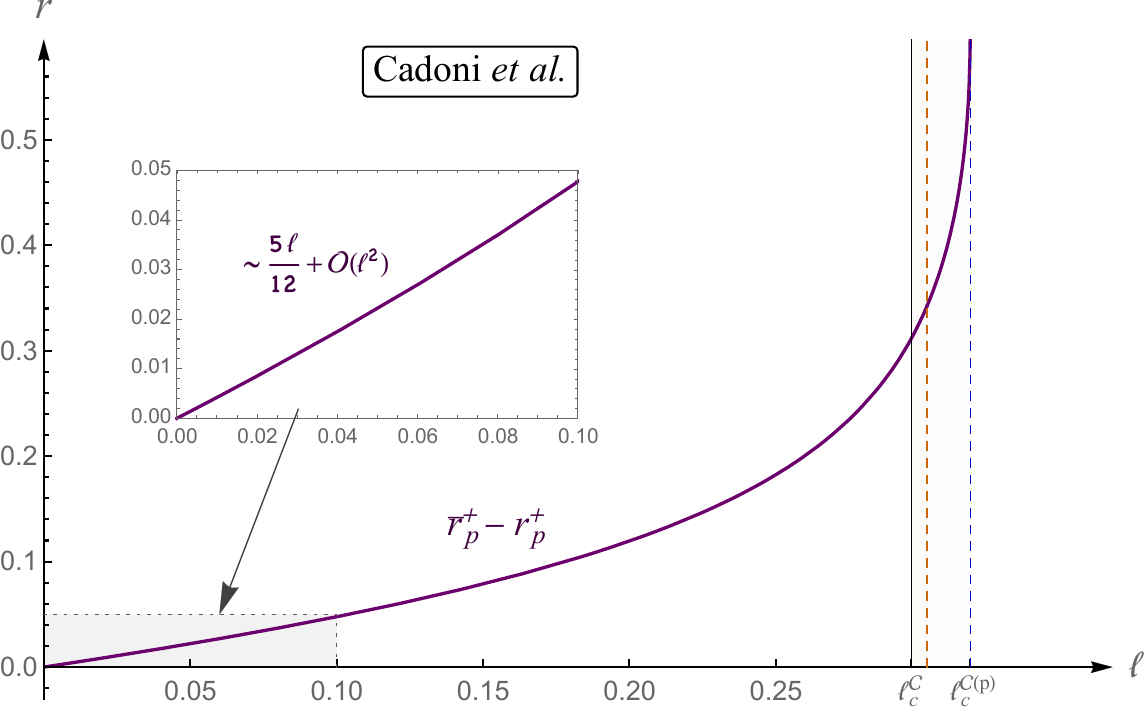}
		\label{subfig:C.Diff}
	}
	\caption{Difference $\bar{r}^{+}_{p} - r^{+}_{p}$ between the outer light ring (LR) in the effective geometry $\bar{r}^{+}_{p}$ (corresponding to the solid red line in Figs.~\ref{subfig:B.RBH}--\ref{subfig:C.RBH}) and the outer light ring in the background geometry $r^{+}_{p}$ (corresponding to the solid blue line in Figs.~\ref{subfig:B.RBH}--\ref{subfig:C.RBH}) for the three nonsingular UCO models considered in Sec.~\ref{sec:RBHs}. For the models proposed by Bardeen and Hayward, the scaling behavior in the regime of small $\ell$ is quadratic, whereas for the Cadoni \textit{et al.}\ model it is linear. Consequently, the outer light rings are more easily distinguished in the Cadoni \textit{el al.}\ model for very small regularization parameter values $\ell$. In all three models, the difference between the outer light rings becomes maximal as the critical light ring length $\ell_c^{(p)}$ of the background geometry (indicated by the vertical dashed blue line) is approached.} 
	\label{fig:Diff}
	\vspace*{-1mm}
\end{figure*}

Table~\ref{tab:critical.lengths} provides an overview of the relevant critical lengths for the three RBH models considered in Sec.~\ref{sec:RBHs}. A universal result is that the outer light ring in the effective geometry (represented by the solid red line in Figs.~\ref{subfig:B.RBH}--\ref{subfig:C.RBH}) is located further away from the nonsingular UCO compared to the outer light ring in the background geometry (represented by the solid blue line in Figs.~\ref{subfig:B.RBH}--\ref{subfig:C.RBH}). Since this result has observational relevance, we illustrate the difference $\bar{r}^{+}_{p} - r^{+}_{p}$ for the three RBH models in Fig.~\ref{fig:Diff}.
	
Another noteworthy feature is that in each geometry (background and effective), there is precisely one observable light ring if the nonsingular UCO is an RBH [$0 < \ell < \ell_c$]. In the interval $\ell_c < \ell < \bar{\ell}_c^{(p)}$ on the other hand, there are three observable light rings in the effective geometry and two in the background geometry as the inner light rings become visible (region shaded in light orange in Figs.~\ref{subfig:B.RBH}--\ref{subfig:C.RBH}). In the interval $\bar{\ell}_c^{(p)} < \ell < \ell_c^{(p)}$, the innermost light rings in the effective geometry have disappeared and only the outermost light ring remains visible, while in the background geometry both the inner and outer light ring are still visible (region shaded in light blue in Figs.~\ref{subfig:B.RBH}--\ref{subfig:C.RBH}). Lastly, in the interval $\ell > \ell^{(p)}_{c}$ only the outer light ring in the effective geometry remains visible. Table~\ref{tab:light.rings} summarizes the number of observable light rings in different domains of the minimal length scale parameter $\ell$ in theories with and without birefringence. 
\enlargethispage{0.05\textheight}

According to a well-established theorem, nonsingular horizonless UCOs have at least two light rings (with one of them being stable) provided that the metric is a regular stationary solution of the Einstein field equations and the spacetime can be continuously deformed into a flat Minkowski spacetime \cite{cbh:17}.\footnote{An alternative approach is considered in Ref.~\cite{d:24}, which derives the stability of the inner light ring of nonsingular horizonless UCOs based on the assumption that their outer light ring has the same properties as that of the Kerr geometry, but without invoking assumptions about the status of energy conditions.}\ However, the effective metric \enlargethispage{0.05\textheight} [Eq.~\eqref{eq:def.eff.metric}] is singular (see App.~\ref{sec:app:singular.eff.geom} for details) and thus the theorem does not apply.\footnote{On the other hand, if the polarization mode that is null in the effective geometry is considered in the background geometry, then the corresponding photon trajectories are no longer null geodesics, and again the theorem does not apply.}\ In fact, we find that theories with birefringence predict an odd number of light rings for $\ell>\ell_c$ (i.e., in the parameter domain where the nonsingular UCO is a horizonless UCO rather than an RBH) and a single light ring remains in the effective geometry for $\ell > \bar{\ell}^{(p)}_{c}$, see Tab.~\ref{tab:light.rings}. This indicates that the number of light rings on its own may not always serve as a definitive indicator in determining the identity of UCOs without knowledge of the underlying theory as both horizonful and horizonless objects may possess the same number of light rings.
\begin{table}[!htpb] 
	\centering
	\resizebox{0.5\textwidth}{!}{ 
		\begin{tabular}{ >{\raggedright\arraybackslash}m{0.20\linewidth}  @{\hskip 0.02\linewidth} >{\centering\arraybackslash}m{0.17\linewidth} @{\hskip 0.02\linewidth} >{\centering\arraybackslash}m{0.17\linewidth} @{\hskip 0.02\linewidth} >{\centering\arraybackslash}m{0.17\linewidth} @{\hskip 0.02\linewidth} >{\centering\arraybackslash}m{0.17\linewidth}} 
			\toprule \toprule
			RBH model & $\ell_{c}$ & $\bar{\ell}^{(p)}_{c}$ & $\ell^{(p)}_{c}$ & $\bar{\ell}^+_\text{min}$ \\ \midrule \\[-2mm]
			Bardeen & $0.7698 M$ & $0.7811 M$ & $0.8587 M$ & $1.0766 M$ \\[3mm]
			Hayward & $0.7698 M$ & $0.7836 M$ & $0.9509 M$ & $1.1004 M$ \\[3mm]
			Cadoni \textit{et al.} & $0.2963 M$ & $0.3016 M$ & $0.3164 M$ & $0.3991 M$ \\[1mm]
			\bottomrule \bottomrule
		\end{tabular}
	}	 
	\vspace*{-1.5mm}
	\caption{Critical lengths $\ell_{c}$, $\bar{\ell}^{(p)}_{c}$, and $\ell^{(p)}_{c}$ at which the horizons, the two innermost light rings in the effective geometry, and the inner and outer light ring in the background geometry merge, respectively. The value $\bar{\ell}^+_\text{min}$ corresponds to the global minimum of the outer light ring in the effective geometry.}
	\label{tab:critical.lengths}
	\vspace*{-5mm}
\end{table} 

\begin{table}[!htpb] 
	\centering 
	\resizebox{0.5\textwidth}{!}{ 
		\begin{tabular}{>{\raggedright\arraybackslash}m{0.19\linewidth} @{\hskip 0.04\linewidth} >{\centering\arraybackslash}m{0.155\linewidth} @{\hskip 0.03\linewidth} >{\centering\arraybackslash}m{0.179\linewidth} @{\hskip 0.03\linewidth} >{\centering\arraybackslash}m{0.198\linewidth} @{\hskip 0.03\linewidth} >{\centering\arraybackslash}m{0.18\linewidth} @{\hskip 0.02\linewidth}} 
			\toprule \toprule
			& $0 \! < \! \ell \! < \! \ell_{c}$ & $\ell_{c} \! < \! \ell \! < \! \bar{\ell}^{(p)}_{c}$ & $\bar{\ell}^{(p)}_{c} \!\! < \! \ell \! < \! \ell^{(p)}_{c}$ & $\ell \! > \! \ell^{(p)}_{c}$ \\ \midrule \\[-2mm]
			\multirow{2}{21.5mm}{Nonsingular \\ UCO type} & \multirow{2}{*}{RBH} & \multirow{2}{*}{Horizonless} & \multirow{2}{*}{Horizonless} & \multirow{2}{*}{Horizonless} \\[6mm]
			\multirow{2}{21.5mm}{LRs without \\ birefringence} & \multirow{2}{*}{$1$} & \multirow{2}{*}{$2$} & \multirow{2}{*}{$2$} & \multirow{2}{*}{$0$} \\[6mm]
			\multirow{2}{21.5mm}{LRs with \\ birefringence} & \multirow{2}{*}{$2$} & \multirow{2}{*}{$5$} & \multirow{2}{*}{$3$} & \multirow{2}{*}{$1$} \\[4mm]
			\bottomrule \bottomrule
		\end{tabular}
	}	 
	\caption{Number of observable light rings (LRs) for different domains of the regularization parameter $\ell$ (corresponding to different types of nonsingular UCOs) in effective theories with and without birefringence. The parameter domains $\ell_{c} \! < \! \ell \! < \! \bar{\ell}^{(p)}_{c}$ and $\bar{\ell}^{(p)}_{c} \!\! < \! \ell \! < \! \ell^{(p)}_{c}$ correspond to the regions shaded in light orange and light blue in Figs.~\ref{subfig:B.RBH}--\ref{subfig:C.RBH} and Figs.~\ref{subfig:B.Diff}--\ref{subfig:C.Diff}, respectively.}
	\label{tab:light.rings}
	\vspace*{-3mm}
\end{table}

In the effective geometry, the outermost light ring of the Hayward model is located the furthest away from the UCO for small values of the regularization parameter $\ell$, followed by those of the Bardeen and Cadoni \textit{et al.}\ model. However, with increasing $\ell$ this behavior changes, and the outermost light ring in the Cadoni \textit{et al.}\ first surpasses that of the Bardeen model and ultimately that of the Hayward model in terms of its distance from the nonsingular UCO, as illustrated by the inset located in the bottom right-hand side corner of Fig.~\ref{subfig:LR.comparison}. The outermost light ring in the effective Hayward geometry is eventually surpassed by that of the Bardeen model as well. These differences in the locations of the outer light ring in the effective geometry for varying $\ell$ are attributable to the different strengths of deformation from the Schwarzschild geometry exhibited by each model.\footnote{A similar feature appears in black hole thermodynamics, where the deviation of the mean-field theory critical ratio also depends on the strength of the deformation from the Schwarzschild geometry \cite{ss:24,s:24}.}\ The different deformation strengths of the three models are also responsible for the different sizes of the intervals $\ell_{c} < \ell < \bar{\ell}^{(p)}_{c}$ and $\bar{\ell}^{(p)}_{c} < \ell < \ell^{(p)}_{c}$ corresponding to the regions shaded in light orange and light blue in Figs.~\ref{subfig:B.RBH}--\ref{subfig:C.RBH} and Figs.~\ref{subfig:B.Diff}--\ref{subfig:C.Diff}, respectively. More precisely: the stronger the deformation from the Schwarzschild geometry, the smaller the size of the critical length scale intervals, as can be verified from the values provided in Tab.~\ref{tab:critical.lengths}. Lastly, we remind the reader that the strength of the deformation is also intimately related to the weak-field limit behavior. Ordering the models we consider in this article in terms of their respective deformation strengths, i.e., $\mathcal{C} > \mathcal{B} > \mathcal{H}$, we observe that the stronger the deformation, the closer the behavior to the Maxwellian weak-field limit, and the smaller the relevant critical length scale intervals.
	
\subsection{Dynamical stability}
The dynamical behavior of light rings is determined by the second-order derivatives of the potential with respect to $r$. For the three UCO models discussed in Sec.~\ref{sec:RBHs}, we find
\begin{align}
	& V''(r^{+}_{p})<0, & V''(r^{-}_{p})>0, \\
	& \bar{V}''(\bar{r}^{+}_{p})<0, \quad \bar{V}''(\bar{r}^{0}_{p})>0, & \bar{V}''(\bar{r}^{-}_{p})<0.
\end{align}
This implies that in the background metric the inner light ring $r^{-}_{p}$ is stable, while the outer light ring $r^{+}_{p}$ is unstable. The outermost light ring $\bar{r}^{+}_{p}$ in the effective geometry exhibits the same instability, but here the innermost light ring $\bar{r}^{-}_{p}$ is also unstable. This is in stark contrast to the non-NED case. Interestingly, the light ring $\bar{r}^{(0)}_{p}$ situated between the inner and outer light ring in the effective geometry is stable. 
	
While even in the effective geometry nonsingular UCOs typically possess a stable light ring leading to well-known spacetime instabilities, this stable light ring is absent beyond a certain minimal length scale, namely $\ell>\bar{\ell}^{(p)}_{c}$. The fact that the outermost light ring $\bar{r}^{+}_{p}$ remains (unlike in the background geometry) suggests that --- if the effective description of nonsingular horizonless UCOs sourced by NED is valid --- they may possess only one unstable light ring, thus presenting a viable alternative to the standard paradigm used to explain observations of astrophysical black hole candidates. Of course, another possibility is that NED theories simply do not provide a viable effective description of nonsingular horizonless UCOs.
	
Lastly, we note that various models for gravitational collapse resulting in nonsingular UCOs violate the null energy condition (NEC), e.g., Friedmann-Robertson-Walker (FRW) collapse models that incorporate repulsive effects to halt the collapse \cite{mt:22,bmm:13} or
thin-shell collapse models where the backreaction of Hawking radiation is described as a trace anomaly \cite{bfk:24}. The emission of Hawking radiation \cite{h:74,h:75} is also known to violate several energy conditions including the NEC, and a violation of the latter is a necessary requirement for the formation of a regular (in the sense of finite curvature scalars) apparent horizon in finite time according to the clock of a distant observer \cite{bhl:18,bmmt:19,mt:21}.

\section{Causality and phase velocities} \label{sec:velocity}
\subsection{Generic expressions}\label{sec:vph:gen}
In this section, we focus on the phase velocities of different photon polarizations and the resulting causal structure of their associated light cones. We begin by computing the phase velocity, utilizing the fact that light propagates along null geodesics. Recall that one photon polarization travels on the background metric, while the other travels on the effective metric. For our analysis, which we once again restrict to the equatorial plane without loss of generality (cf.\ Sec.~\ref{subsec:LR.mp}), we proceed with a propagation wavevector $k_{\mu}$ of the form
\begin{align}
	k_{\mu}=\left(-\omega,\sqrt{\tensor{\sg}{_r_r}}|\vec{k}|\cos{\eta},0,\sqrt{\tensor{\sg}{_\phi_\phi}}|\vec{k}|\sin{\eta}\right),
	\label{eq:k}
\end{align}
where $|\vec{k}|=\sqrt{\tensor{\sg}{^i^j}k_{i}k_{j}}$, with indices $i,j$ running from $1$ to $3$, and $\eta\in\left[0,\pi\right]$ representing an angle introduced to conveniently specify the propagation direction in what follows. Radial light rays are described by $\eta=0$, whereas $\eta=\pi/2$ for circular trajectories. The phase velocity is defined as $\omega/|\vec{k}|$. For the background geometry, the null condition $\tensor{\sg}{^\mu^\nu} k_{\mu} k_{\nu} = 0$ [cf.\ Eq.~\eqref{eq:bkg.nc}] and Eq.~\eqref{eq:k} lead to
\begin{align}
	v_{\text{ph}} = \sqrt{-\tensor{\sg}{_t_t}} \stackrel{\eqref{eq:metric}}{=} \sqrt{f(r)} .
	\label{eq:bkg.vph}
\end{align}
Since this expression is independent of $\eta$, the phase velocity is the same in any direction of the photon propagation. Similarly, using Eq.~\eqref{eq:def.eff.metric} and Eq.~\eqref{eq:eff.nc} for the effective metric, we find
\begin{align}
	\tensor{{\sg}}{^\mu^\nu} k_{\mu} k_{\nu} - 4 \frac{\mathcal{L}_{\mathcal{F}\mathcal{F}}}{\mathcal{L}_\mathcal{F}} \tensor{\sg}{_\rho_\sigma} \left( \tensor{F}{^\mu^\rho}k_{\mu} \right) \left( \tensor{F}{^\nu^\sigma}k_{\nu} \right) = 0 ,
	\label{eq:dispersion}
\end{align}
and the phase velocity is given by
\begin{align}
	\bar{v}_{\text{ph}}(\eta) = \sqrt{f(r) \left( 1 + \frac{2 \mathcal{F} \mathcal{L}_{\mathcal{F}\mathcal{F}}}{\mathcal{L}_\mathcal{F}} \sin^2{\eta} \right)} . 
	\label{eq:eff.vph}
\end{align}
Unlike the phase velocity in the background geometry given by Eq.~\eqref{eq:bkg.vph}, the phase velocity in the effective geometry Eq.~\eqref{eq:eff.vph} depends on the photon's direction of travel. Consequently, due to the fact that one photon polarization mode moves according to the background geometry while the other moves according to the effective geometry, their respective phase velocities will differ, signifying the presence of birefringence. It is worth emphasizing that this phenomenon does not affect radial light rays described by a propagation vector with $\eta=0$ [cf.\ Eq.~\eqref{eq:k}]. In this case, $v_{\text{ph}} = \bar{v}_{\text{ph}}(0)$, and thus radial light rays propagate with the same phase velocity.
	
As discussed in Sec.~\ref{subsec:B.RBH} and Sec.~\ref{subsec:H.RBH}, the effective metric tensor components of the Bardeen and Hayward models do not conform to the proper Schwarzschild limit in the case of vanishing minimal length. This property prevents a direct comparison of the phase velocity in these spacetimes to that of the singular Schwarzschild geometry. In the subsequent analysis, we therefore mainly focus on the model by Cadoni \textit{et al.}\ [cf.\ Sec.~\ref{subsec:C.RBH}]. 
	
\subsection{Phase velocities in the Cadoni \textit{et al.}\ model}\label{sec:vph:C}
\begin{figure}[!htbp]
	\includegraphics[width=.95\linewidth]{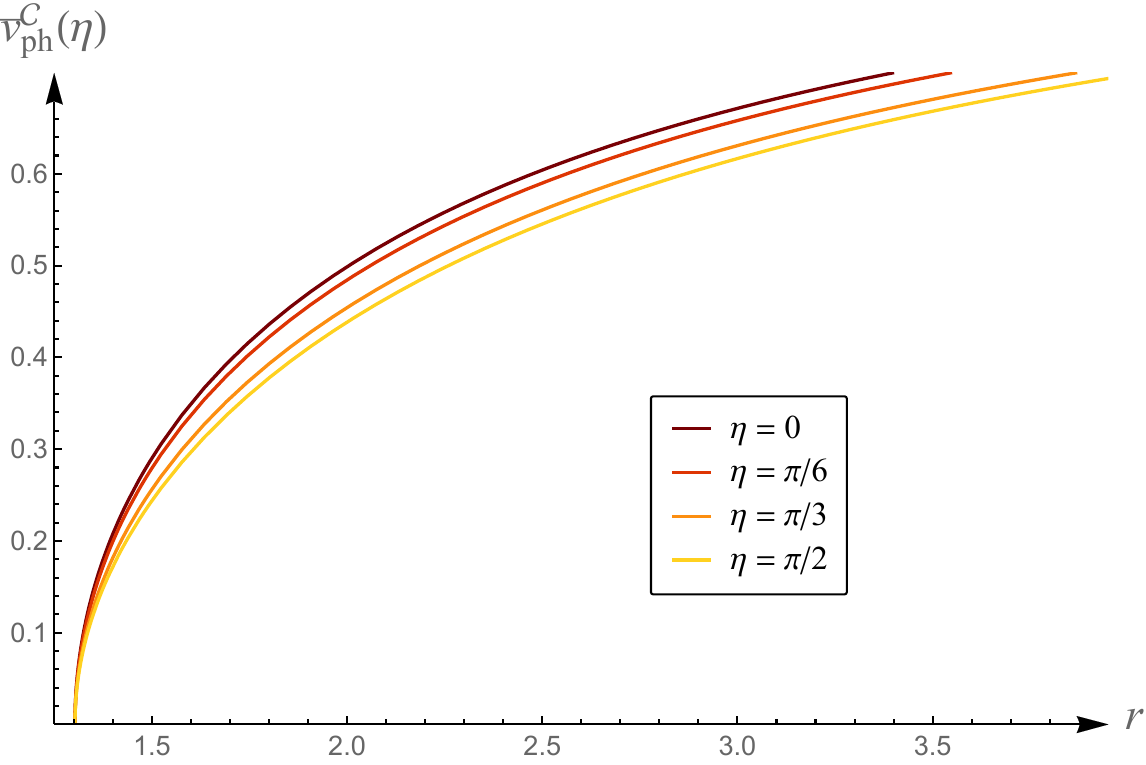}
	\vspace*{-1.1mm}
	\caption{Illustration of the dependence of the phase velocity $\bar{v}_\text{ph}^\mathcal{C}(\eta)$ in the effective metric of the Cadoni \textit{et al.}\ model [Eq.~\eqref{eq:C.RBH.eff.vph}] on the propagation angle $\eta \in [0,\pi]$ for $M=1$ and $\ell = 0.2$. From the darkest to the lightest hue, the lines correspond to the values $\eta = \lbrace 0, \frac{\pi}{6}, \frac{\pi}{3}, \frac{\pi}{2} \rbrace$.} 
	\label{fig:EtaDependence}
	\vspace*{-3.5mm}
\end{figure}

In this model, the first and second derivative of the Lagrangian density $\mathcal{L}_{\mathcal{C}}(\mathcal{F})$ [Eq.~\eqref{eq:C.RBH.L}] with respect to $\mathcal{F}$ are given as functions of the radial coordinate $r$ by
\begin{align}
	\mathcal{L}'_{\mathcal{C}}(\mathcal{F}) = \frac{12 r^5}{(r+\ell)^{5}} , \quad \mathcal{L}''_{\mathcal{C}}(\mathcal{F}) = -\frac{15r^9}{M(r+\ell)^6} ,
	\label{eq:C.UCO.LF.LFF}
\end{align}
where we have substituted the corresponding values of the parameters $\alpha$ and $Q_{m}$ given by Eq.~\eqref{eq:C.RBH.alpha.Qm} after taking the derivatives. Via Eq.~\eqref{eq:bkg.vph} the phase velocity in the background geometry of the Cadoni \textit{et al.}\ model is given by
\begin{align}
	v^{(\mathcal{C})}_{\text{ph}} = \sqrt{f_{\mathcal{C}}(r)} .
	\label{eq:C.RBH.bkg.vph}
\end{align}
Using Eqs.~\eqref{eq:A.Qm.F}, \eqref{eq:C.RBH.alpha.Qm}, and \eqref{eq:C.UCO.LF.LFF} in Eq.~\eqref{eq:eff.vph} it follows that the phase velocity in the effective geometry of the Cadoni \textit{et al.}\ model is given by 
\begin{align}
	\bar{v}^{(\mathcal{C})}_{\text{ph}}(\eta) = \sqrt{f_{\mathcal{C}}(r) \left(1-\frac{5 \ell}{2(r+\ell)} \sin^{2}\eta \right)} .
	\label{eq:C.RBH.eff.vph}
\end{align} 
Fig.~\ref{fig:EtaDependence} depicts its dependence on the propagation angle $\eta$.
	
The following two comments are in order: First, as $r\rightarrow \infty$, the influence of the magnetic charge $Q_m$ or, equivalently [via Eq.~\eqref{eq:C.RBH.alpha.Qm}], the minimal length scale $\ell$, diminishes, as one would expect intuitively, and the effective phase velocity reduces to that of the background geometry (which coincides with that of the Schwarzschild geometry in this limit), demonstrating the consistency of the calculation. Second, in the limit of vanishing minimal length, the Cadoni \textit{et al.}\ model reduces to the Schwarzschild limit for every direction of motion, in contrast to the models by Bardeen and Hayward (see Sec.~\ref{sec:vph:HB}).
	
For radial trajectories described by $\eta=0$, the phase velocity of the Cadoni \textit{et al.}\ model surpasses that in the Schwarzschild geometry, which can be attributed to the presence of a minimal length scale and consequently the absence of a singularity. The velocities coincide only in the asymptotic regime where the presence of a nonvanishing $\ell$ becomes insignificant.
	
\begin{figure}[!htbp]
	\includegraphics[width=.95\linewidth]{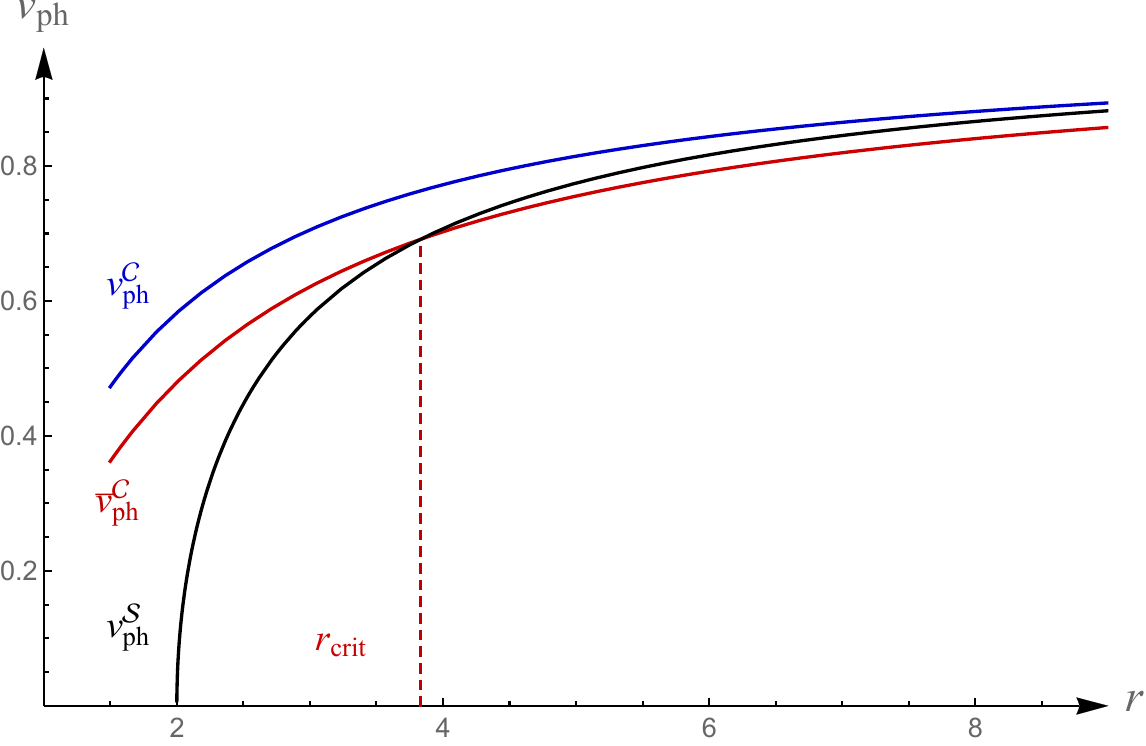}
	\vspace*{-2.2mm}
	\caption{Comparison of the phase velocities in the background $v_\text{ph}^\mathcal{C}$ (blue) and effective $\bar{v}_\text{ph}^\mathcal{C}$ (red) geometries of the nonsingular Cadoni \textit{et al.}\ model at its critical length $\ell = \ell^\mathcal{C}_c$ to the phase velocity $v_\text{ph}^\mathcal{S}$ in the singular Schwarzschild geometry for $M=1$ and $\eta=\pi/2$. For these parameter choices, the critical radius which signifies the point beyond which the phase velocity in the Schwarzschild geometry exceeds that in the effective geometry is given by $r_{\text{crit}} = 3.8313$.} 
	\label{fig:PhaseVelocities}
\end{figure}

Nonradial light rays with $\eta \neq 0$ propagating in the effective geometry exhibit an intriguing behavior illustrated in Fig.~\ref{fig:PhaseVelocities}: such rays possess a higher phase velocity compared to those of the Schwarzschild geometry, but only up to a certain critical radius $r_{\text{crit}}$ which increases with increasing $M$ and decreasing $\ell$. Beyond this critical radius, their phase velocity is smaller compared to the Schwarzschild case.\footnote{We emphasize that, when we are concerned with circular light trajectories, the phase velocity as a function of $r$ is only meaningful if such trajectories exist. As can be seen in Fig.~\ref{fig:Horizons.LRs}, light rings may be located at distances greater than the location of the Schwarzschild light ring, and therefore it is still meaningful to consider their phase velocity in this regime of $r$.}\ In addition, it is evident from Eqs.~\eqref{eq:C.RBH.bkg.vph} and \eqref{eq:C.RBH.eff.vph} that 
\begin{align}
	v^{(\mathcal{C})}_{\text{ph}}>\bar{v}^{(\mathcal{C})}_{\text{ph}},
\end{align}	       	
indicating that the wavefronts associated with the background geometry always propagate faster than those of the effective geometry. Moreover, they also always propagate faster than those in the Schwarzschild geometry, cf.\ Fig.~\ref{fig:PhaseVelocities}. 
	
Nevertheless, the phase velocity never becomes superluminal for any propagation direction, a property that is intimately related to the fact that $\mathcal{L}_{\mathcal{C}}''(\mathcal{F})<0$ for the Cadoni \textit{et al.}\ model. As demonstrated in Sec.~\ref{sec:vph:HB}, this feature is not shared by the Bardeen and Hayward models. The significance of the signature of $\mathcal{L}''(\mathcal{F})$ becomes clear when examining the light cone structure. It is important to note that the wavevector $k_{\mu}$ is a 1-form (covariant vector), while the photon momentum, i.e., the tangent vector to the photon trajectory, is a contravariant vector \cite{s:02}. In the linear Maxwell theory, this distinction is unnecessary, but it is necessary in NED theories with both a background and an effective metric. To accurately determine the nature of a null trajectory in the effective geometry with respect to the background geometry, we must start with the covariant form of the effective metric and the contravariant momentum vector: 
\begin{align}
	\bar{\sg}_{\mu\nu}k^{\mu}k^{\nu} = 0 .
\end{align}
Using the covariant components of the effective metric given by Eqs.~\eqref{eq:eff.metric.tt.rr}, \eqref{eq:eff.metric.thetatheta}, and \eqref{eq:eff.metric.phiphi}, we have 
\begin{align}
	k^{2} = (-\tensor{\sg}{_t_t})\frac{2 \mathcal{F} \mathcal{L}_{\mathcal{F}\mathcal{F}}}{\mathcal{L}_{\mathcal{F}}} \left(k^{t}\right)^2,
	\label{eq:k^2-cov}
\end{align}
where $k^2=\tensor{\sg}{_\mu_\nu}k^{\mu}k^{\nu}$. Since both $-\tensor{\sg}{_t_t}$ and $\left(k^{t}\right)^2$ are positive and $\mathcal{L}_{\mathcal{F}}>0$ for all UCO models considered in this article (see Sec.~\ref{sec:vph:HB}), the nature of a trajectory that is null in the effective geometry propagating within the background geometry is determined by the sign of $\mathcal{L}_{\mathcal{F}\mathcal{F}}$. Specifically, for $\mathcal{L}_{\mathcal{F}\mathcal{F}}<0$ the vector $k^{\mu}$ that is null in the effective geometry is timelike in the background geometry, thereby preserving the causal character of the theory. 
	
Alternatively, one can start with the null condition parsed in terms of the contravariant effective metric components,
\begin{align}
	\bar{\sg}^{\mu\nu}k_{\mu}k_{\nu} = 0 .
\end{align}
An explicit calculation then leads to the expression 
\begin{align}
	k^{2} = - \tensor{\sg}{^\phi^\phi} \frac{2 \mathcal{F} \mathcal{L}_{\mathcal{F}\mathcal{F}}}{\mathcal{L}_\mathcal{F}} \left(k_{\phi}\right)^2 .
\end{align}
While this appears to contradict the argumentation following Eq.~\eqref{eq:k^2-cov}, in this case the vector $k_{\mu}$ is spacelike in the background metric for $\mathcal{L}_{\mathcal{F}\mathcal{F}}<0$, and the light cone of the background geometry lies entirely within the light cone of the effective geometry, indicating superluminal propagation velocities. The observation that the nesting of light cones reverses when transitioning from the contravariant metric to the covariant metric is a well-known result \cite{gh:01}. Hence, these two approaches of addressing causality are indeed equivalent, and it is ultimately the signature of $\mathcal{L}_{\mathcal{F}\mathcal{F}}$ that plays a crucial role.
	
\subsection{Phase velocities in the Bardeen and Hayward models} \label{sec:vph:HB}
Following the methodology of Sec.~\ref{sec:vph:gen}, we calculate the phase velocities in the UCO models by Bardeen [Sec.~\ref{subsec:B.RBH}] and Hayward [Sec.~\ref{subsec:H.RBH}] based on Eqs.~\eqref{eq:bkg.vph} and \eqref{eq:eff.vph} for the background and effective geometries, respectively. 
	
The derivatives of the NED Bardeen Lagrangian density [Eq.~\eqref{eq:B.RBH.L}] with respect to $\mathcal{F}$ are given by
\begin{align}
	\mathcal{L}'_{\mathcal{B}}(\mathcal{F}) &= \frac{15 \ell r^6}{(r^2 + \ell^2)^{7/2}} , \\ \mathcal{L}''_{\mathcal{B}}(\mathcal{F}) &= \frac{15 r^{10} (r^2 - 6 \ell^2)}{4M(r^2 + \ell^2)^{9/2}} , 
	\label{eq:Lf-B}
\end{align}
where we have substituted the parameters $\alpha$ and $Q_{m}$ from Eq.~\eqref{eq:B.RBH.alpha.Qm}. The corresponding phase velocities in the background and effective geometry are 
\begin{align}
	v^{\mathcal{(B)}}_{\text{ph}} &= \sqrt{f_{\mathcal{B}}(r)} , 
	\label{eq:vph-back-B} \\
	\bar{v}^{\mathcal{(B)}}_{\text{ph}} &= \sqrt{{f_{\mathcal{B}}(r)} \left( 1 + \frac{\left(r^2-6\ell^2\right)}{2\left(r^2+\ell^2\right)} \sin^2 \eta \right)} ,
	\label{eq:vph-eff-B}
\end{align}
respectively. For the NED Hayward Lagrangian density [Eq.~\eqref{eq:H.RBH.L}], the analogous expressions are given by
\begin{align}
	\mathcal{L}'_{\mathcal{H}}(\mathcal{F}) &= \frac{18 (2 M \ell^2)^{2/3} r^{7}}{(r^3 + 2 M \ell^2)^3},
	\label{eq:Lf-H1} \\
	\mathcal{L}''_{\mathcal{H}}(\mathcal{F}) &= \frac{9 \cdot 2^{1/3} \ell^{2/3} (r^3 - 7 M \ell^2) r^{11}}{M^{2/3}(r^3 + 2 M \ell^2)^{4}} ,
	\label{eq:Lf-H2}
\end{align}
where we have substituted the parameters $\alpha$ and $Q_{m}$ from Eq.~\eqref{eq:H.RBH.alpha.Qm}. The phase velocities in the background and effective geometry are given by
\begin{align}
	v^{(\mathcal{H})}_{\text{ph}} &= \sqrt{f_{\mathcal{H}}(r)} , 
	\label{eq:vph-back-H} \\
	\bar{v}^{(\mathcal{H})}_{\text{ph}} &= \sqrt{f_{\mathcal{H}}(r) \left(1 + \frac{r^3 - 7 M \ell^2}{r^3 + 2 M \ell^2} \sin^2 \eta \right)} . 
	\label{eq:vph-eff-H}
\end{align}
Based on the relations given in Eqs.~\eqref{eq:vph-back-B} and \eqref{eq:vph-back-H}, we can conclude that the Schwarzschild phase velocity is recovered in the limit $\ell \rightarrow 0$ only for radial null rays described by $\eta=0$. This demonstrates once again that for radial null rays the phase velocities in the background and effective geometries coincide, and therefore they do not experience birefringence. Nonradially propagating null rays on the other hand experience birefringence, as illustrated in Figs.~\ref{fig:EtaDependence} and \ref{fig:PhaseVelocities} for the Cadoni \textit{et al.}\ model.
	
Lastly, it is worth pointing out a distinctive behavior that occurs in the models by Bardeen and Hayward but is absent in the Cadoni \textit{et al.}\ model. In the latter, $\mathcal{L}''_{\mathcal{C}}(\mathcal{F})<0$, ensuring a phase velocity less than one describing a propagation that is causal. However, this is not the case for polarized light moving in the effective geometries of the Bardeen and Hayward models. Specifically, there exists a radial distance $r_{\star}$ that leads to a signature change in $\mathcal{L}''_{\mathcal{B}}(\mathcal{F})$ and $\mathcal{L}''_{\mathcal{H}}(\mathcal{F})$. From Eqs.~\eqref{eq:Lf-B} and \eqref{eq:Lf-H2}, these critical radii are identified as
\begin{align}
	r_{\star}^{(\mathcal{B})} = \sqrt{6} \ell , \quad r_{\star}^{(\mathcal{H})} = \left(7 M \ell^2\right)^{1/3} ,
\end{align}
respectively. According to Eq.~\eqref{eq:k^2-cov}, this implies that a null trajectory in the effective metric behaves spacelike in the background metric for $r>r_{\star}^{(\mathcal{B})}$ and $r>r_{\star}^{(\mathcal{H})}$, indicating regions of superluminal propagation (provided that physically relevant trajectories exist in these regions, which depends on the value of the minimal length scale $\ell$). To verify if superluminal signalling is possible in practice requires an examination of the group velocity of the associated light pulses \cite{b:60}, which may be discussed further in future works.

\section{Discussion and conclusions} \label{sec:discussion.conclusions}
NED theories possess many intriguing features and the possibility of regularizing black hole geometries by coupling them to gravity remains an interesting proposal. Their nonlinearity results in the violation of the superposition principle, causing propagating light rays to be affected by electromagnetic background fields \cite{plm:18}. This phenomenon, known as ``light-by-light scattering'', has been experimentally observed \cite{b*:97}. Vacuum birefringence is another interesting effect, according to which different photon polarizations propagate with different phase velocities. Therefore, if a light ray propagates through a region pervaded by a strong electromagnetic field, it will effectively be separated into two distinct rays. Although this phenomenon has not yet been definitively observed in nature, there is compelling evidence from studies of light rays passing by magnetars --- neutron stars with extremely powerful magnetic fields ranging from $10^{9}$ to $10^{11}$ T \cite{m*:17}.
	
Due to the difficulties in maintaining regularity at the center while simultaneously conforming to the Maxwell weak-field limit [Eq.~\eqref{eq:weak-field.limit}] inherent to solutions with an electric charge (cf.\ Sec~\ref{sec:magn.sol}), our analysis focuses on the geometries of purely magnetic nonsingular UCOs sourced by NED. We examine their observational properties such as light ring signatures and phase velocities. Our analysis illustrates that the phenomenon of birefringence may manifest itself through the presence of additional light rings surrounding the nonsingular UCO (cf.\ Sec.~\ref{sec:LR}, Fig.~\ref{fig:Horizons.LRs}). On the other hand, the number of light rings on its own is insufficient to distinguish RBHs from nonsingular horizonless UCOs as both models may possess either one or two light rings depending on the minimal length scale and the presence or absence of birefringence (cf.\ Sec.~\ref{sec:LR}, Tab.~\ref{tab:light.rings}). 

Interestingly, our analysis reveals that there are narrow intervals in the minimal length scale parameter, namely $\ell_{c}<\ell<\ell^{(p)}_{c}$ (corresponding to the union of the regions shaded in light orange and light blue in Figs.~\ref{subfig:B.RBH}--\ref{subfig:C.RBH} and Figs.~\ref{subfig:B.Diff}--\ref{subfig:C.Diff}) and $\ell_{c}<\ell<\bar{\ell}^{(p)}_{c}$ (corresponding to the region shaded in light orange in Figs.~\ref{subfig:B.RBH}--\ref{subfig:C.RBH} and Figs.~\ref{subfig:B.Diff}--\ref{subfig:C.Diff}), in which the inner light ring in the background geometry and the two innermost light rings in the effective geometry, respectively, become visible to external observers. Another notable result is that the outer light ring in the effective geometry persists beyond the critical light ring length $\ell^{(p)}_{c}$ of the background geometry, as illustrated in Fig.~\ref{fig:Horizons.LRs}.
	
While our procedure is generic based on the general form of the NED Lagrangian density proposed by Fan and Wang [Ref.~\cite{fw:16}, Eq.~(26)], we explicitly consider three popular nonsingular UCO models characterized by the strength of their respective deviations from the singular Schwarzschild geometry. Only one of these models, namely that considered by Cadoni \textit{et al.}\ in Ref.~\cite{c*:23}, possesses an effective NED metric that exhibits the desired behavior in the limit of vanishing minimal length and in the asymptotic regime (Sec.~\ref{subsec:C.RBH}). For this particular model, the minimal length parameter is bounded from above by $\ell \lesssim 0.47M$ via observations of the S2 star orbiting Sagittarius A$^*$, the black hole candidate located at the center of the Milky Way Galaxy \cite{c*:23}. However, since the critical light ring lengths of this model sit far below this value (cf.\ last row in Tab.~\ref{tab:critical.lengths}), the current bound is insufficient to 
\begin{enumerate}[label=\roman*)]
	\item distinguish between an RBH and a nonsingular horizonless UCO;
	\item either exclude or confirm the presence of birefringence;
\end{enumerate}
based on the number of light rings surrounding the UCO. 
	
Testing the existence of additional light rings is only possible if the resolution achieved in our astrophysical observations is sufficient to distinguish them. To determine the required resolution for nonsingular UCOs sourced by NED, we compute the light ring locations and compare their differences. A universal result is that the outermost light ring in the effective geometry is always situated at larger radii compared to the outer light ring of the background geometry. As illustrated in Fig.~\ref{fig:Diff}, the separation between these two light rings increases with the minimal length scale, reaching its maximum at $\ell^{(p)}_c$, the critical light ring length where the inner and outer light ring of the background geometry merge and disappear. While the minimal length scale need not be Planckian, it is typically assumed to be small. It is therefore useful to examine how the light ring separation behaves in this regime. We find that in the model by Cadoni \textit{et al.}, the separation scales with $\sim \ell$, whereas in the Bardeen and Hayward models it scales with $\sim \ell^2$. This characteristic behavior can be attributed to the stronger deformation of the Cadoni \textit{et al.}\ model from the Schwarzschild geometry compared to the other two models. As a result, detecting both light rings individually is a more feasible task for the Cadoni \textit{et al.}\ model at small length scales.
	
A different interpretation for the light rings in NED geometries is offered in Ref.~\cite{ist:23}, according to which photons move exclusively in what we refer to as the effective geometry. Complementary analyses related to our interpretation according to which one of the two effective geometries coincides with the background geometry are given in Refs.~\cite{hkm:24,gb:24} for electrically charged solutions in the class of regularized Maxwell (RegMax) theories and dyonic solutions in ModMax theories, respectively. The light rings, shadows, and gravitational lensing effects of the electrically charged Dymnikova RBH \cite{d:92} with an NED source are studied in Ref.~\cite{dlcc:23}.
	
In addition to light ring signatures, we analyzed the phase velocities for different photon polarizations and examined the corresponding causal structure for the three nonsingular UCO models considered in this article. Our results indicate that there are no acausal spacetime regions when the NED theory adheres to the Maxwell weak-field limit as in the Cadoni \textit{et al.}\ model. On the other hand, in models where this is not the case such as those proposed by Bardeen and Hayward, the light cone of the background geometry is fully contained within the light cone of the effective geometry, which indicates the presence of spacetime regions where superluminal motion is possible, provided that the relevant trajectories exist. 
	
The limiting behavior of the Cadoni \textit{et al.}\ model allows for comparisons to the Schwarzschild geometry. We find that the phase velocity in the background geometry exceeds that of the effective geometry as well as that of the Schwarzschild geometry. Interestingly, the phase velocity in the effective geometry exceeds the Schwarzschild phase velocity only up to some critical radius, beyond which it is smaller, as depicted in Fig.~\ref{fig:PhaseVelocities}. Measuring these phase velocities in experiments could provide an alternative way of establishing bounds on the minimal length scale parameter.
	
Astrophysical observations of black hole candidates are typically modeled using the Kerr paradigm since realistic UCOs are expected to possess angular momentum and rotate. In particular, the emission of gravitational waves requires at least a mass quadrupole structure and thus cannot be modeled in spherically symmetric settings. On the other hand, a recent study indicates that spherically symmetric solutions in semiclassical gravity can mimick signatures of axially symmetric geometries \cite{st:24}.\ Nonetheless, to comprehensively test the validity of NED theories as an effective description of nonsingular UCOs requires an extension to axial symmetry. However, attempts to provide such an extension using the Newman-Janis formalism have not been successful thus far. Fully analytic solutions remain unknown, and only solutions for the case of slowly rotating objects are currently available \cite{hkst:24,kts:22}.

\section*{Acknowledgements}
We would like to thank Pedro Cunha, Carlos Herdeiro, Yasha Neiman, and Daniel Terno for useful discussions and helpful comments. SM is supported by the Quantum Gravity Unit of the Okinawa Institute of Science and Technology (OIST). IS is supported by an International Macquarie University Research Excellence Scholarship (IMQRES).

\appendix
\renewcommand{\thesection}{\Alph{section}}
\renewcommand{\thesubsection}{\Alph{section}.\arabic{subsection}}
\renewcommand{\thesubsubsection}{\Alph{section}.\arabic{subsection}.\arabic{subsubsection}}
\section{Birefringence} \label{sec:app:birefringence}
Following the argumentation of Ref.~\cite{bb:70}, we re-derive the phenomenon of birefringence for convenience. As in Sec.~\ref{sec:NED}, we restrict our considerations to NED Langrangian densities $\mathcal{L}(\mathcal{F},\mathcal{G}) \equiv \mathcal{L}(\mathcal{F})$. The derivation is based on the assumption that the electromagnetic field $\tensor{\tilde{F}}{_\mu_\nu}$ of the UCO is much stronger than that of photons $\tensor{\Phi}{_\mu_\nu}$. For simplicity, we work in Minkowski spacetime here, but an analogous derivation can be performed in generic curved spacetimes by considering covariant derivatives instead of partial derivatives in what follows. The total electromagnetic field is given by
\begin{align}
	\tensor{F}{_\mu_\nu} = \tensor{\tilde{F}}{_\mu_\nu} + \tensor{\Phi}{_\mu_\nu} .
\end{align}
Defining
\begin{align}
	\tensor{h}{^\mu^\nu} \defeq \mathcal{L}_\mathcal{F} \tensor{F}{^\mu^\nu} ,
	\label{eq:def:h}
\end{align}
the leading terms in the series expansion about $\tensor{\Phi}{_\mu_\nu}$ are
\begin{align}
	\tensor{h}{^\mu^\nu} = \tensor{h}{^\mu^\nu} \big\vert_{\Phi=0} + \frac{\partial \tensor{h}{^\mu^\nu}}{\partial \tensor{F}{_\rho_\sigma}} \bigg\vert_{F=\tilde{F}} \tensor{\Phi}{_\rho_\sigma} + \mathcal{O}(\Phi^2) ,
	\label{eq:h.leadingTerms}
\end{align}
where indices have been omitted in the $\vert_\bullet$ subscripts and order $\mathcal{O}(\bullet)$ here and in what follows to avoid clutter. Substituting Eq.~\eqref{eq:h.leadingTerms} into Eq.~\eqref{eq:def:h} results in the linearized equations
\begin{align}
	\partial_\mu \tensor{h}{^\mu^\nu} = \frac{\partial \tensor{h}{^\mu^\nu}}{\partial \tensor{F}{_\rho_\sigma}} \bigg\vert_{F=\tilde{F}} \partial_\mu \tensor{\Phi}{_\rho_\sigma} + \mathcal{O}(\Phi) = 0 .
	\label{eq:h.linearised}
\end{align}
We are looking for solutions $\tensor{\Phi}{_\rho_\sigma}$ of the form
\begin{align}
	\tensor{\Phi}{_\rho_\sigma}(x) = \tensor{\epsilon}{_\rho_\sigma}(k) e^{- ikx} ,
	\label{eq:Phi.field}
\end{align}
where
\begin{align}
	\tensor{\epsilon}{_\rho_\sigma} \defeq k_\rho \epsilon_\sigma(k) - k_\sigma \epsilon_\rho(k)
	\label{eq:def:antisym.p.tensor}
\end{align}
denotes the antisymmetric polarization tensor, $k_\rho$ the propagation vector, and $\epsilon_\rho$ the polarization vector \cite{w:book:72}. Substituting the partial derivative
\begin{align}
	\partial_\mu \tensor{\Phi}{_\rho_\sigma} = - i k_\mu \tensor{\epsilon}{_\rho_\sigma}(k) e^{-ikx}
\end{align}
into Eq.~\eqref{eq:h.linearised} results in
\begin{align}
	\frac{\partial \tensor{h}{^\mu^\nu}}{\partial \tensor{F}{_\rho_\sigma}} \bigg\vert_{F=\tilde{F}} \left(-i k_\mu \tensor{\epsilon}{_\rho_\sigma}(k) e^{-ikx} \right) = 0 ,
\end{align}
and using the definition of the antisymmetric polarization tensor Eq.~\eqref{eq:def:antisym.p.tensor}, we have
\begin{align}
	\frac{\partial \tensor{h}{^\mu^\nu}}{\partial \tensor{F}{_\rho_\sigma}} \bigg\vert_{F=\tilde{F}} k_\mu \left(k_\rho \epsilon_\sigma - k_\sigma \epsilon_\rho \right) = 0 .
\end{align}
Finally, using the antisymmetry of the electromagnetic field tensor $\tensor{F}{_\rho_\sigma}$ [cf.\ Eq.~\eqref{eq:def:Ftensor}], we arrive at the relation
\begin{align}
	\frac{\partial \tensor{h}{^\mu^\nu}}{\partial \tensor{F}{_\rho_\sigma}} \bigg\vert_{F=\tilde{F}} k_\mu k_\rho \epsilon_\sigma = 0 .
	\label{eq:pd.h.k.k.eps}
\end{align}
The partial derivative is calculated as
\begin{align}
	\frac{\partial \tensor{h}{^\mu^\nu}}{\partial \tensor{F}{_\rho_\sigma}} \stackrel{\eqref{eq:def:h}}{=} \frac{\partial}{\partial \tensor{F}{_\rho_\sigma}} \left( \mathcal{L}_\mathcal{F} \tensor{F}{^\mu^\nu} \right) = \mathcal{L}_{\mathcal{F}\mathcal{F}} \frac{\partial \mathcal{F}}{\partial \tensor{F}{_\rho_\sigma}} \tensor{F}{^\mu^\nu} + \mathcal{L}_\mathcal{F} \frac{\partial \tensor{F}{^\mu^\nu}}{\partial \tensor{F}{_\rho_\sigma}} ,
	\label{eq:pd.h}
\end{align}
where
\begin{align}
	\frac{\partial \tensor{F}{^\mu^\nu}}{\partial \tensor{F}{_\rho_\sigma}} = \frac{1}{2} \left( \tensor{\eta}{^\mu^\rho} \tensor{\eta}{^\nu^\sigma} - \tensor{\eta}{^\nu^\rho} \tensor{\eta}{^\mu^\sigma} \right)
	\label{eq:pd.F.F}
\end{align}
due to the antisymmetry property [cf.\ Eq.~\eqref{eq:def:Ftensor}]. The derivative of the field strength $\mathcal{F}$ with respect to $\tensor{F}{_\rho_\sigma}$ is calculated using this relation and results in
\begin{align}
	\begin{aligned}
		\frac{\partial \mathcal{F}}{\partial \tensor{F}{_\rho_\sigma}} &= \frac{\partial}{\partial \tensor{F}{_\rho_\sigma}} \left( \tensor{F}{_\alpha_\beta} \tensor{F}{^\alpha^\beta} \right) \\
		&= \frac{\partial}{\partial \tensor{F}{_\rho_\sigma}} \left( \tensor{\eta}{_\alpha_\chi} \tensor{\eta}{_\beta_\lambda} \tensor{F}{^\alpha^\beta} \tensor{F}{^\chi^\lambda} \right) = 2 \tensor{F}{^\rho^\sigma} .
	\end{aligned}
	\label{eq:pd.cF.F}
\end{align}
Substitution of Eqs.~\eqref{eq:pd.F.F} and \eqref{eq:pd.cF.F} into Eq.~\eqref{eq:pd.h} and subsequently into Eq.~\eqref{eq:pd.h.k.k.eps} results in
\begin{align}
	2 \mathcal{L}_{\mathcal{F}\mathcal{F}} \left( \tensor{F}{^\rho^\sigma} k_\rho \right) \left( \tensor{F}{^\mu^\nu} k_\mu \right) \epsilon_\sigma + \frac{1}{2} \mathcal{L}_\mathcal{F} \left( \tensor{\eta}{^\nu^\sigma} k^2 - k^\nu k^\sigma \right) \epsilon_\sigma = 0 .
	\label{eq:pd.h.two.terms}
\end{align}
It is useful to define the four-vectors
\begin{align}
	a^\mu \defeq \tensor{F}{^\mu^\nu} k_\nu , \quad \tensor{^\star a}{^\mu} \defeq \tensor{^\star F}{^\mu^\nu} k_\nu , \quad b^\mu \defeq \tensor{F}{^\mu^\nu} a_\nu .
\end{align}
Equation~\eqref{eq:pd.h.two.terms} can then be rewritten as
\begin{align}
	\left[\mathcal{L}_\mathcal{F}\left(\tensor{\eta}{^\nu^\sigma} k^2 - k^\nu k^\sigma \right) + 4 \mathcal{L}_{\mathcal{F}\mathcal{F}} a^\nu a^\sigma \right] \epsilon_\sigma = 0 .
	\label{eq:pd.h.bracket}
\end{align}
Defining
\begin{align}
	\tensor{M}{^\nu^\sigma} \defeq \mathcal{L}_\mathcal{F}\left(\tensor{\eta}{^\nu^\sigma} k^2 - k^\nu k^\sigma \right) + 4 \mathcal{L}_{\mathcal{F}\mathcal{F}} a^\nu a^\sigma ,
\end{align}
Eq.~\eqref{eq:pd.h.bracket} is given by
\begin{align}
	\tensor{M}{^\nu^\sigma} \epsilon_\sigma = 0 .
	\label{eq:M.eps}
\end{align}
To determine the polarization vector $\epsilon_\sigma$, we express it in terms of four linearly independent vectors $a^\mu$, $^\star a^\mu$, $k^\mu$, and $b^\mu$, i.e.,
\begin{align}
	\epsilon_\sigma = c_1 a_\sigma + c_2 {^\star a_\sigma} + c_3 k_\sigma + c_4 b_\sigma .
\end{align}
Equation~\eqref{eq:M.eps} is then given by
\begin{align}
	c_1 \tensor{M}{^\nu^\sigma} a_\sigma + c_2 \tensor{M}{^\nu^\sigma} ^\star a_\sigma + c_3 \tensor{M}{^\nu^\sigma} k_\sigma + c_4 \tensor{M}{^\nu^\sigma} b_\sigma = 0 .
	\label{eq:c.M.eps}
\end{align}
Using $k^\sigma a_\sigma = k^\sigma \tensor{F}{_\sigma_\alpha} k^\alpha = 0$ (due to antisymmetry), the first term is given by
\begin{align}
	\tensor{M}{^\nu^\sigma} a_\sigma = \left( \mathcal{L}_\mathcal{F} k^2 + 4 \mathcal{L}_{\mathcal{F}\mathcal{F}} a^2 \right) a^\nu ,
	\label{eq:c1term} 
\end{align}	
where $a^2 \defeq a_\mu a^\mu$. Similarly, $k^\sigma {^\star a}_\sigma = k^\sigma \tensor{^\star F}{_\sigma_\nu} k^\nu = 0$ (due to antisymmetry), and the second term is given by
\begin{align}
	\tensor{M}{^\nu^\sigma} ^\star a_\sigma = \mathcal{L}_\mathcal{F} k^2 {^\star a}^\nu + 4 \mathcal{L}_{\mathcal{F}\mathcal{F}} \left(a^\sigma {^\star a}_\sigma \right) a^\nu .
	\label{eq:c2term} 
\end{align}
Analogously, the third and fourth term are given by
\begin{align}
	\tensor{M}{^\nu^\sigma} k_\sigma &= \mathcal{L}_\mathcal{F} \left(k^2 k^\nu - k^2 k^\nu \right) + 4 \mathcal{L}_{\mathcal{F}\mathcal{F}} \left(a^\sigma k_\sigma\right) a^\nu = 0 , 
	\label{eq:c3term} \\
	\tensor{M}{^\nu^\sigma} b_\sigma &= \mathcal{L}_\mathcal{F} k^2 b^\nu - \mathcal{L}_\mathcal{F} \left(k^\sigma b_\sigma\right) k^\nu ,
	\label{eq:c4term}
\end{align}
respectively, where we have used $a^\sigma k_\sigma = 0$ and $a^\sigma b_\sigma = a^\sigma \tensor{F}{_\sigma_\alpha} a^\alpha = 0$ (due to anitsymmetry). Substitution of Eqs.~\eqref{eq:c1term}--\eqref{eq:c4term} into Eq.~\eqref{eq:c.M.eps} yields
\begin{align}
	& \left[ c_1 \left( \mathcal{L}_\mathcal{F} k^2 + 4 \mathcal{L}_{\mathcal{F}\mathcal{F}} a^2 \right) + c_2 4 \mathcal{L}_{\mathcal{F}\mathcal{F}} \left(a^\sigma {^\star a}_\sigma \right) \right] a^\nu \label{eq:coefficientEQ} \\
	& \; + c_2 \left( \mathcal{L}_\mathcal{F} k^2 \right) {^\star a}^\nu - c_4 \left[\mathcal{L}_\mathcal{F} \left( k^\sigma b_\sigma \right) \right] k^\nu + c_4 \left( \mathcal{L}_\mathcal{F} k^2 \right) b^\nu = 0 \nonumber .
\end{align}
Since the four-vectors $a^\mu$, $^\star a^\mu$, $k^\mu$, and $b^\mu$ are linearly independent, the vanishing of $\tensor{M}{^\nu^\sigma} \epsilon_\sigma$ implies that the coeffcients must vanish. We note that
\begin{align}
	k^\sigma b_\sigma = k^\sigma \tensor{F}{_\sigma_\alpha} a^\alpha = - \left(\tensor{F}{_\alpha_\sigma} k^\sigma\right) a^\alpha = - a_\alpha a^\alpha = - a^2 ,
\end{align}
and
\begin{align}
	a^\sigma {^\star a}_\sigma &= \tensor{F}{^\sigma^\beta} k_\beta \tensor{^\star F}{_\sigma_\alpha} k^\alpha = \tensor{\eta}{^\alpha^\lambda} \left(- \mathcal{G} \eta^\beta_\alpha \right) k_\beta k_\lambda \nonumber \\
	&= - \mathcal{G} \tensor{\eta}{^\beta^\lambda} k_\sigma k_\lambda = - \mathcal{G} k^2 ,
\end{align}
where we have used the property $\tensor{F}{^\sigma^\beta} \tensor{^\star F}{_\sigma_\alpha} = - \mathcal{G} \eta^\beta_\alpha$ with $\mathcal{G}$ as defined in Eq.~\eqref{eq:NEDinvariants}.

Since the last two terms in Eq.~\eqref{eq:coefficientEQ} cannot vanish simultaneously unless $c_4=0$, we are left with a system of equations for the coefficients $c_1$ and $c_2$.\footnote{Note that the coefficient $c_3$ does not contribute to the polarization as evident from Eq.~\eqref{eq:coefficientEQ}. This may also be seen from Eqs.~\eqref{eq:Phi.field} and \eqref{eq:def:antisym.p.tensor}: if $\epsilon^\mu \propto k^\mu$, then the polarization tensor $\tensor{\epsilon}{_\rho_\sigma}$ and its corresponding field vanishes. Physically, this implies that the polarization vector is always perpendicular to the propagation direction. Hence no part of the polarization vector lies in the propagation direction and thus $c_3$ does not contribute.}\ For $k^2=0$, we have
\begin{align}
	c_1 \left(4 \mathcal{L}_{\mathcal{F}\mathcal{F}} a^2 \right) = 0 ,
\end{align}
and an arbitrary $c_2$, which necessarily leads to $c_1=0$, and thus the polarization vector is
\begin{align}
	\epsilon^\mu = c_2 {^\star a^\mu} .
	\label{eq:bkg.pol.vec}
\end{align}
This corresponds to the photon polarization mode that moves on null geodesics in the background geometry. For $k^2 \neq 0$ on the other hand, we find
\begin{align}
	c_1 \left( \mathcal{L}_\mathcal{F} k^2 + 4 \mathcal{L}_{\mathcal{F}\mathcal{F}} a^2 \right) = 0 .
\end{align}
To find a nontrivial (i.e., $c_1 \neq 0$) solution to the eigenvalue problem $\tensor{M}{^\nu^\sigma} \epsilon_\sigma = 0$, we solve
\begin{align}
	\mathcal{L}_\mathcal{F} k^2 + 4 \mathcal{L}_{\mathcal{F}\mathcal{F}} a^2 = 0 \;\; \Rightarrow \;\; k^2 + \frac{4 \mathcal{L}_{\mathcal{F}\mathcal{F}}}{\mathcal{L}_\mathcal{F}} a^2 = 0 .
\end{align}
Using the definition of the four-vector $a^\mu$, this equation can be rewritten as
\begin{align}
	& k^2 + \frac{4 \mathcal{L}_{\mathcal{F}\mathcal{F}}}{\mathcal{L}_\mathcal{F}} \tensor{\eta}{_\mu_\nu} \tensor{F}{^\mu^\alpha} k_\alpha \tensor{F}{^\nu^\beta} k_\beta = 0 \\
	\Rightarrow \; \; & \tensor{\eta}{^\alpha^\beta} k_\alpha k_\beta + \frac{4 \mathcal{L}_{\mathcal{F}\mathcal{F}}}{\mathcal{L}_\mathcal{F}} \tensor{\eta}{_\mu_\nu} \tensor{F}{^\mu^\alpha} \tensor{F}{^\nu^\beta} k_\alpha k_\beta = 0 \\
	\Rightarrow \; \; & \!\! \left( \tensor{\eta}{^\alpha^\beta} + \frac{4 \mathcal{L}_{\mathcal{F}\mathcal{F}}}{\mathcal{L}_\mathcal{F}} \tensor{\eta}{_\mu_\nu} \tensor{F}{^\mu^\alpha} \tensor{F}{^\nu^\beta} \right) k_\alpha k_\beta = 0 .
\end{align}
Consequently, the effective metric tensor is given by
\begin{align}
	\tensor{\bar{\sg}}{^\alpha^\beta} = \tensor{\eta}{^\alpha^\beta} + \frac{4 \mathcal{L}_{\mathcal{F}\mathcal{F}}}{\mathcal{L}_\mathcal{F}} \tensor{\eta}{_\mu_\nu} \tensor{F}{^\mu^\alpha} \tensor{F}{^\nu^\beta} .
\end{align}
Using antisymmetry of the electromagnetic field tensor $\tensor{F}{^\nu^\beta} = - \tensor{F}{^\beta^\nu}$ and contracting $\tensor{\eta}{_\mu_\nu} \tensor{F}{^\beta^\nu}$ results in
\begin{align}
	\tensor{\bar{\sg}}{^\alpha^\beta} = \tensor{\eta}{^\alpha^\beta} - \frac{4 \mathcal{L}_{\mathcal{F}\mathcal{F}}}{\mathcal{L}_\mathcal{F}} \tensor{F}{^\alpha_\mu} \tensor{F}{^\mu^\beta} ,
\end{align}
which is equivalent to Eq.~\eqref{eq:def.eff.metric}. Thus the polarization vector
\begin{align}
	\epsilon^\mu = c_1 a^\mu
	\label{eq:eff.pol.vec}
\end{align}
corresponds to the photon polarization mode that moves on null geodesics in the effective geometry.

\subsection{Examples}
In what follows, we consider circular trajectories. As in Sec.~\ref{sec:LR}, we restrict our considerations to the equatorial plane without loss of generality, and thus the wavevector can be written as $k_\mu = (\omega,0,0,k_\phi)$.

\subsubsection{Magnetic solutions}
For purely magnetic solutions [$Q_e \! = \! 0 \Rightarrow \! \tensor{F}{_t_r} \! = \! 0$], the polarization vector that is null in the background geometry [Eq.~\eqref{eq:bkg.pol.vec}] is given explicitly by
\begin{align}
	\epsilon^\mu &= c_2 {^\star a}^\mu = \frac{c_2}{2} \tensor{^\star F}{^\mu^\nu} k_\nu = \frac{c_2}{2} \tensor{\varepsilon}{^\mu^\nu^\rho^\sigma} \tensor{F}{_\rho_\sigma} k_\nu \nonumber \\
	&= c_2 \tensor{\varepsilon}{^\mu^\nu^\theta^\phi} \tensor{F}{_\theta_\phi} k_\nu = (0,c_2 \omega \tensor{F}{_\theta_\phi},0,0) ,
\end{align}
corresponding to a polarization in the radial direction and a propagation in the $\phi$ direction.

For the polarization vector that is null in the effective geometry [Eq.~\eqref{eq:eff.pol.vec}], we have
\begin{align}
	\bar{\epsilon}^\mu &= c_1 a^\mu = c_1 \tensor{F}{^\mu^\nu} k_\nu = c_1 \tensor{F}{^\theta^\phi} k_\phi \nonumber \\
	&= (0,0,c_1 \tensor{F}{^\theta^\phi} k_\phi,0) ,
\end{align}
corresponding to a polarization in the $\theta$ direction and a propagation in the $\phi$ direction.

\subsubsection{Electric solutions}
For purely electric solutions [$Q_m \! = 0 \Rightarrow \tensor{F}{_\theta_\phi} \! = 0$], we find
\begin{align}
	\epsilon^\mu &= \frac{c_2}{2} \tensor{\varepsilon}{^\mu^\nu^\rho^\sigma} \tensor{F}{_\rho_\sigma} k_\nu = c_2 \tensor{\varepsilon}{^\mu^\nu^t^r} \tensor{F}{_t_r} k_\nu \nonumber \\
	&= (0,0,c_2 \tensor{F}{_t_r} k_\phi,0) ,
\end{align}
and
\begin{equation}
	\bar{\epsilon}^\mu = c_1 \tensor{F}{^\mu^\nu} k_\nu = (0,c_1 \omega \tensor{F}{^t^r},0,0) ,
\end{equation}
respectively.

\section[Singularity of the effective metric]{\\ Singularity of the effective metric} \label{sec:app:singular.eff.geom}
The line element described by the covariant effective metric tensor components of Eqs.~\eqref{eq:eff.metric.tt.rr}--\eqref{eq:eff.metric.phiphi} is given by
\begin{align}
	ds^2 = -f_{i}(r)dt^2+\frac{dr^2}{f_{i}(r)}+\frac{r^2}{\bar{h}_{i}(r)}d\Omega^2,
	\label{eq:ds.eff}
\end{align}
where the subscript $i \in \lbrace \mathcal{B},\mathcal{H},\mathcal{C} \rbrace$ labels functions associated with the Bardeen [Eq.~\eqref{eq:RBH.B.f}], Hayward [Eq.~\eqref{eq:RBH.H.f}], and Cadoni \textit{et al.} [Eq.~\eqref{eq:RBH.C.f}] model, respectively, and
\begin{align}
	\bar{h}_{i}(r) \defeq 1 + \frac{2 \mathcal{F} \mathcal{L}_{\mathcal{F}\mathcal{F}}}{\mathcal{L}_{\mathcal{F}}},
\end{align}	
with $\mathcal{L}(\mathcal{F})$ given by Eqs.~\eqref{eq:B.RBH.L}, \eqref{eq:H.RBH.L}, and \eqref{eq:C.RBH.L}, respectively. Explicit evaluation yields [cf.\ Eqs.~\eqref{eq:B.RBH.eff.metric.thetatheta}--\eqref{eq:B.RBH.eff.metric.phiphi}, \eqref{eq:H.RBH.eff.metric.thetatheta}--\eqref{eq:H.RBH.eff.metric.phiphi}, and \eqref{eq:C.RBH.eff.metric.thetatheta}--\eqref{eq:C.RBH.eff.metric.phiphi}]
\begin{align}
	\bar{h}_{\mathcal{B}}(r) &= \frac{3}{2}-\frac{7\ell^2}{2(r^2+\ell^2)}, \\
	\bar{h}_{\mathcal{H}}(r) &= 2-\frac{9M\ell^2}{r^3+2M\ell^2}, \\
	\bar{h}_{\mathcal{C}}(r) &= 1-\frac{5 \ell}{2(r+\ell)}.
\end{align}
Based on these explicit expressions, the equation $\bar{h}_{i}(r)=0$ admits real positive solutions for $r$, indicating a divergence of the effective metric tensor components $\tensor{\bar{\sg}}{_\theta_\theta}$ and $\tensor{\bar{\sg}}{_\phi_\phi}$. The existence of a curvature singularity may also be confirmed via examination of the Ricci (or Kretschmann) scalar corresponding to Eq.~\eqref{eq:ds.eff}, i.e.,
\begin{align}
	\begin{aligned}
		\bar{\mathcal{R}}_{i}(r) &= \frac{2 \bar{h}_{i} \! \left(2 \bar{h}^2_{i} + 2 r^2 f'_{i} \bar{h}'_{i} - r \bar{h}_{i} \left(4 f'_{i} + r f''_{i} \right) \right)}{2 r^2 \bar{h}_{i}^2} \\
		& \quad + \frac{f_{i} \big( 4 r \bar{h}_{i} \left(3 \bar{h}'_{i} + r \bar{h}''_{i} \right) - 4 \bar{h}^2_{i} - 7 r^2 (\bar{h}'_{i})^2 \big)}{2 r^2 \bar{h}^2_{i}} ,
	\end{aligned}
	\label{eq:Ricci-gen}
\end{align}
where primes denote differentiation with respect to $r$ and explicit dependencies of the functions $f_{i}(r)$ and $\bar{h}_{i}(r)$ on $r$ have been omitted for the sake of simplicity. Evaluating Eq.~\eqref{eq:Ricci-gen} for each of the three models yields
\begin{align}
	\bar{\mathcal{R}}_{\mathcal{B}}(r) &= \frac{\mathcal{A}_{\mathcal{B}}(r)}{(r^2+\ell^2)^{7/2}(3 r^3 - 4 r \ell^2)^2}, \\
	\bar{\mathcal{R}}_{\mathcal{H}}(r) &= \frac{\mathcal{A}_{\mathcal{H}}(r)}{2(r^3+2M\ell^2)^3(2 r^4 - 5 M r \ell^2)^2}, \\
	\bar{\mathcal{R}}_{\mathcal{C}}(r) &= \frac{\mathcal{A}_{\mathcal{C}}(r)}{2 r^2 (2 r - 3 \ell)^2 (r+\ell)^5},
\end{align}
which diverge at $r_\mathcal{B} = \frac{2 \ell}{\sqrt{3}}$, $r_\mathcal{H} = (\frac{5}{2})^{1/3} M^{1/3} \ell^{2/3}$, and $r_\mathcal{C} = \frac{3 \ell}{2}$, respectively, and the numerators are given explicitly by

\vspace*{-4mm}
\begin{widetext}
	\begin{align}
		\mathcal{A}_{\mathcal{B}}(r) &= (r^2+\ell^2)^{3/2} \left( 9 r^8 - 69 r^6 \ell^2 - 268 r^4 \ell^4 - 384 r^2 \ell^6 - 96 \ell^8 \right) \nonumber \\
		& \qquad \quad + 2M \left( 57 r^8 \ell^2 + 250 r^6 \ell^4 + 336 r^4 \ell^6 + 192 r^2 \ell^8 \right) , \\[4mm]
		\mathcal{A}_{\mathcal{H}}(r) &= 16 r^{15} + 8M(84M-43r) r^{11} \ell^2 + M^2 (3726M-2879r) r^8 \ell^4 + 2 M^3 (5040M-4517r) r^5 \ell^6 \nonumber \\
		& \qquad \quad + 80M^4 (60M-137r) r^2 \ell^8 - 2800 M^5 \ell^{10}, \\[4mm]
		\mathcal{A}_{\mathcal{C}}(r) &= \ell^2 \left( - 5 (r+\ell)^3 (7 r^2 + 30 r \ell + 18 \ell^2) + 2 M r^2 (71 r^2 + 12 r \ell + 216 \ell^2) \right) .
	\end{align}
\end{widetext}

\end{document}